\documentclass[12pt]{article}
\usepackage{rotating}
\usepackage{amsmath}
\usepackage{amsthm}
\usepackage{amsfonts}
\usepackage[dvips]{epsfig}
\usepackage[utf8]{inputenc}
\usepackage{graphics,graphicx}
\usepackage{float}
\usepackage{amssymb}
\usepackage{cancel}
\usepackage{pstricks} 
\usepackage{lscape}
\usepackage{cite}
\usepackage{color}
\newcommand{\beq}{\begin{equation}}
\newcommand{\eeq}{\end{equation}}
\newcommand{\ga}{\lower.7ex\hbox{$\;\stackrel{\textstyle>}{\sim}\;$}}
\newcommand{\la}{\lower.7ex\hbox{$\;\stackrel{\textstyle<}{\sim}\;$}}

\begin{document}

\def\thefootnote{\fnsymbol{footnote}}

\begin{flushright}
{\tt KCL-PH-TH/2014-40}, {\tt LCTS/2014-40}, {\tt CERN-PH-TH/2014-195}  \\
\end{flushright}

\vspace{0.3cm}
\begin{center}
{\bf {\Large Exploring CP Violation in the MSSM}}
\vspace {0.3cm}
\end{center}


\begin{center}{\large
{\bf A.~Arbey}$^{a,b}$
{\bf J.~Ellis}$^{c,b}$,
{\bf R.~M.~Godbole}$^d$
and
{\bf F.~Mahmoudi}$^{a,b,}$\footnote{Also Institut Universitaire de France, 103 boulevard Saint-Michel, 75005 Paris, France}
}
\end{center}

\begin{center}
{\em $^a$Universit{\' e} de Lyon, Universit{\' e} Lyon 1, F-69622 Villeurbanne Cedex, France;\\
Centre de Recherche Astrophysique de Lyon, CNRS, UMR 5574, Saint-Genis Laval Cedex, F-69561, France;
Ecole Normale Sup{\' e}rieure de Lyon, France
}\\[0.2cm]
{\em $^b$Theory Division, CERN, CH-1211 Geneva 23,
  Switzerland}\\[0.2cm]
{\em $^c$Theoretical Particle Physics and Cosmology Group, Department of
  Physics, King's~College~London, London WC2R 2LS, United Kingdom}\\[0.2cm]
{\em $^d$Centre for High Energy Physics, Indian Institute of Science, Bangalore, 560012, India}\\
\end{center}

\bigskip

\centerline{\bf ABSTRACT}

\noindent  
{\small We explore the prospects for observing CP violation in the minimal supersymmetric extension of the
Standard Model (MSSM) with six CP-violating parameters, three gaugino mass phases and three phases in
trilinear soft supersymmetry-breaking parameters, using the {\tt CPsuperH} code combined with a geometric approach
to maximise CP-violating observables subject to the experimental upper bounds on electric dipole moments.
We also implement CP-conserving constraints from Higgs physics, flavour physics and the upper limits on
the cosmological dark matter density and spin-independent scattering.
We study possible values of observables within the constrained MSSM (CMSSM), the non-universal Higgs
model (NUHM), the CPX scenario and a variant of the phenomenological MSSM (pMSSM). We find values of
the CP-violating asymmetry $A_{CP}$ in $b \to s \gamma$ decay that may be as large as 3\%, so future
measurements of $A_{CP}$ may provide independent information about CP violation in the MSSM. We find that
CP-violating MSSM contributions to the $B_s$ meson mass mixing term $\Delta M_{B_s}$ are in general
below the present upper limit, which is dominated by theoretical uncertainties. If these could be
reduced, $\Delta M_{B_s}$ could also provide an interesting and complementary constraint on the six
CP-violating MSSM phases, enabling them all to be determined experimentally, in principle.
We also find that CP violation in the $h_{2,3} \tau^+ \tau^-$ and $h_{2,3} {\bar t} t$ couplings can be quite large, and
so may offer interesting prospects for future $pp$, $e^+ e^-$, $\mu^+ \mu^-$ and $\gamma \gamma$  colliders.}

\renewcommand{\thefootnote}{\arabic{footnote}}
\setcounter{footnote}{0}

\section{Introduction}

The minimal supersymmetric extension of the Standard Model (MSSM) contains many
possible sources of CP violation beyond the Kobayashi-Maskawa phase of the Standard Model
and the strong CP phase. These additional sources of CP violation arise from the soft
supersymmetry-breaking terms in the low-energy effective Lagrangian, and include phases in
the gaugino masses, the trilinear scalar couplings and the sfermion mass matrices. However,
the Kobayashi-Maskawa phase accounts very well for the CP-violating effects seen in the
$K^0$ system and in $B$ meson decays, and no other violations of CP have been observed
despite, for example, sensitive experimental searches for electric dipole moments (EDMs). Thus, it might be
tempting to suggest that the extra MSSM sources of CP violation are absent. On the other hand,
experimental upper limits still allow considerable scope for additional CP-violating effects in,
for example, $B^0_s$ mixing, and some additional source of CP violation is needed
to explain the cosmological baryon asymmetry, which might be due to these MSSM phases.
For these reasons, there have been many studies of possible MSSM CP-violating effects in
experimental observables, and powerful phenomenological tools have been developed for
calculating these effects.

In view of the success of the Cabibbo-Kobayashi-Maskawa (CKM) model in describing flavour
mixing and CP violation in the quark sector, it is often assumed that the strong CP phase
is negligibly small for some reason, and that flavour and CP violation for squarks is generated by
the CKM mixing in the quark sector, the hypothesis of minimal flavour violation (MFV). However,
even in this case there remain several additional sources of CP violation in the MSSM, namely
the phases in the gaugino masses and the trilinear couplings. One is thus led to consider the
maximally CP-violating, minimal flavour-violating (MCPMFV) model that contains six CP-violating
phases beyond the Kobayashi-Maskawa phase: three phases $\Phi_{1,2,3}$ in the masses of the
U(1), SU(2) and SU(3) gauginos, and 3 phases $\Phi_{A_{t,b,\tau}}$ in the trilinear soft supersymmetry-breaking
couplings $A_{t,b,\tau}$ of the third-generation stop, sbottom and stau sfermions, respectively~\footnote{We
assume that the strong CP phase is negligible, and also neglect the phases in the trilinear couplings of the
sfermions in the first and second generations, which
are much less important for phenomenology.}. In this study, we allow the six CP-violating phases
to vary independently in all the scenarios considered. Predictions of the MCPMFV scenario for
CP-violating observables such as the CP-violating asymmetry in $b \to s \gamma$ decay, $A_{CP}$,
the CP-violating phase in $B_s$ mixing, $\phi_s$, and EDMs have been considered in~\cite{Ellis:2007kb,Ellis:2010xm},
and possibilities for probing these CP-violating phases through the polarisation of third-generation fermions, $t$ and $\tau$,  produced in the decays of the corresponding sfermions have also been explored~\cite{Gajdosik:2004ed}.

It might be thought that the MSSM phases $\Phi_{1,2,3,t,b,\tau}$ must necessarily be small,
in view of the stringent upper limits on several EDMs shown in Table~\ref{tab:EDMs}.
However, this is not necessarily the case, since there are four main independent EDM
constraints on what is, {\it a priori}, a 6-dimensional space of CP-violating MSSM phases,
so there are in principle `blind directions' corresponding to combinations of phases that do
not `see' the EDM constraints. In principle, individual phases could be large along these directions,
as discussed in~\cite{Olive:2005ru} for example,
and could have significant effects on other CP-violating observables such as $A_{CP}$ and $\phi_s$~\footnote{We
note in passing that there are also well-motivated supersymmetric models in which the phases are naturally
small, so that the EDM bounds are not very constraining, see~\cite{Ellis:2014tea} for example.}.

\begin{table}[!t]
\begin{center}
\begin{tabular}{|c|c|c|}
 \hline
 EDM & Upper limit (e.cm) & Reference\\
 \hline\hline
 Thallium & $1.3\times10^{-24}$ & \cite{Regan:2002ta}\\
 \hline
 Mercury & $3.5\times10^{-29}$ & \cite{Griffith:2009zz}\\
 \hline
 Neutron & $4.7\times10^{-26}$ &  \cite{Baker:2006ts}\\
  \hline
 Thorium monoxide & $1.1\times10^{-28}$ & \cite{Baron:2013eja} \\
 \hline
\end{tabular}
\caption{\it The 95\% CL upper limits on EDMs used as constraints in this study. The present experimental
upper bound on the EDM of the muon, $1.9\times10^{-19}$~e.cm~\cite{Bennett:2008dy}, provides only
a very weak constraint that is not competitive with the other EDM constraints in the models discussed here.
\label{tab:EDMs}}
\end{center}
\end{table}

A brute force way to study this possibility would be to sample randomly the 6-dimensional space of
CP-violating MSSM phases, but this is not the most efficient procedure to explore the possible
magnitudes of CP-violating effects in the MSSM. If one wishes to generate a
large sample of parameter sets that respect other phenomenological constraints such as those
from the flavour, Higgs and dark matter sectors, one would prefer to optimise the search for MSSM
scenarios with maximal CP violation. A geometric approach to this problem was proposed in~\cite{Ellis:2010xm},
and used to analyse the impacts of three EDM constraints in certain specific benchmark MSSM scenarios.

In this paper we adapt and extend this geometric approach to study systematically the possible
magnitudes of CP-violating effects in light of the updated EDM constraints shown in Table~\ref{tab:EDMs}.
The inclusion of a fourth EDM constraint requires a slight extension of the analysis based on three EDMs
made in~\cite{Ellis:2010xm}, as we discuss in Section~2. Also, the geometric approach was originally
formulated as a linear expansion around the CP-conserving limit, whereas we are interested in the
largest possible values of the CP-violating phases. Accordingly, here we extend the approach using
an iterative procedure, finding an initial `blind direction' as in~\cite{Ellis:2010xm}, then choosing a CP-violating
point along that direction with non-zero phases, and then repeating the geometrical optimisation in a new
linear approximation around this CP-violating point, as also discussed in Section~2. In Section~3 we then apply
the geometric approach to four variants of the MSSM, a best-fit 
scenario~\cite{Buchmueller:2013rsa,Buchmueller:2014yva} within the constrained MSSM (CMSSM) in which the
soft supersymmetry-breaking parameters are constrained to be universal at the GUT scale (apart from
the CP-violating phases), a generalisation of this model in which the soft supersymmetry-breaking
contributions to the two Higgs doublet masses are allowed to vary independently (NUHM2), a version of the CPX
scenario defined in~\cite{Carena:2000ks} that is
modified to be in agreement with the LHC results, and the phenomenological MSSM (pMSSM)~\cite{Djouadi:1998di},
in which extrapolation to
the GUT scale is ignored and universality is not imposed~\footnote{Partial results in the case of CMSSM
were presented in~\cite{Brooijmans:2014eja}.}.
In each case, in addition to the EDM constraints in Table~\ref{tab:EDMs}, we also consider the
relevant constraints from flavour physics, from the measured properties of the known Higgs boson
and searches for other MSSM Higgs bosons, and upper limits on the cosmological density of
dark matter and the direct detection of dark matter via scattering on nuclei. 

We focus, in particular, on four possible signatures of MSSM CP violation: the possibility
that there might be another neutral Higgs boson lighter than the one already discovered by ATLAS and CMS,
the CP-violating asymmetry in $b \to s \gamma$ decay, $A_{CP}$, and the non-Standard-Model
contribution to the $B_s$ meson mixing parameter, $\Delta M_{B_s}$, and CP-violating couplings
of the heavier neutral Higgs bosons.
We find that, although a neutral MSSM Higgs boson lighter than that discovered would be consistent with the
EDM constraints, it is excluded by the available limits on other Higgs bosons, notably the absence of
a light charged Higgs boson. Secondly, we find that values of $A_{CP} \lesssim 3$\% are allowed by
the EDMs and other constraints in some of the MSSM scenarios
studied. This opens up the possibility that $A_{CP}$ could be significantly larger than in the Standard Model,
providing a signature of CP-violating MSSM. Conversely, if a non-zero value of $A_{CP}$ were not to
be found in future experiments, this could provide a constraint on CP violation in the MSSM that is
independent of, and complementary to, those from EDMs. Thirdly, in the case of $\Delta M_{B_s}$,
we find that it could also provide an independent constraint on the CP-violating MSSM if the
theoretical uncertainties could be reduced, thereby enabling in principle a complete determination
of all the phases for fixed values of the CP-conserving MSSM parameters. Fourthly,
we also find that CP violation in the $h_{2,3} \tau^+ \tau^-$ and $h_{2,3} {\bar t} t$ couplings can be quite large,
and may offer interesting prospects for future $pp$, $e^+ e^-$ and $\mu^+ \mu^-$ experiments.

\section{Method}
\label{sec:method}

In this Section we outline our approach to sampling the parameter spaces of MSSM scenarios while respecting
the four EDM constraints in Table~\ref{tab:EDMs}. Since the EDM constraints are quite strong,
they effectively reduce the dimensionality of any MSSM scenario by four. The challenge is to sample efficiently this subspace of codimension four, so as to assess how large any other CP-violating observable may be.
Moreover, the thorium monoxide EDM constraint on the electron EDM is now so strong that we have designed a new method to
sample effectively the parameter space.
We do this by adapting and extending the geometric approach proposed in~\cite{Ellis:2010xm}. In the first
Subsection we discuss how the approach may be modified to take into account four EDM constraints,
in the following Subsection we describe an extension of the analysis beyond the small-phase approximation,
and in the third Subsection we summarise our sampling algorithm.

\subsection{Geometric Approach to Maximising a CP-Violating Observable with Four EDM Constraints}

Initially, we consider the four EDMs $E^{a,b,c,d}$ of Table~\ref{tab:EDMs} in the small-phase approximation~\footnote{We
use Latin indices $i, j, ...$ for the EDMs, and Greek indices $\alpha, \beta, ...$ for the CP-violating phases.},
where
\begin{equation}
E^i \; \simeq \; {\mathbf \Phi}.{\mathbf E}^i \, ,
\label{smallangle}
\end{equation}
with ${\bf \Phi} \equiv \Phi_\alpha = \Phi_{1,2,3,t,b,\tau}$ and ${\mathbf E}^i \equiv \partial E^i/\partial {\mathbf \Phi}$
(i.e., $E^i_\alpha \equiv \partial E^i/\partial \Phi_\alpha$).
The ${\mathbf \Phi}$ subspace of codimension four is spanned by the following quadruple exterior product:
\begin{equation}
 A_{\alpha\beta\gamma\delta} \; = \; E^a_{[ \alpha} \, E^b_\beta \, E^c_\gamma \, E^d_{\delta ]} \, 
\label{subspace}
\end{equation}
where the symbols $[ ... ]$ denote antisymmetrisation of the enclosed indices. This subspace is a
2-dimensional plane, as in the simple example in Section~2.1 of~\cite{Ellis:2010xm}. We now consider some
CP-violating observable $O$ whose dependence on the phases $\Phi_\alpha$ is given in the
small-phase approximation by ${\mathbf O} \equiv \partial O/\partial {\mathbf \Phi}$
(i.e., $O_\alpha \equiv \partial O/\partial \Phi_\alpha$). One can then define the vector
\begin{equation}
 B_\mu \; \equiv \; \epsilon_{\mu\nu\lambda\rho\sigma\tau} \, O_\nu \, E^a_\lambda \, E^b_\rho \, E^c_\sigma \, E^d_\tau
\end{equation}
that characterises a direction in the space of CP-violating phases where there is no contribution to
the observable $O$, nor to the EDMs. The EDM-free direction that optimises $O$ is clearly
orthogonal to $B_\mu$ as well as to the EDM vectors $E^{a,b,c,d}_\alpha$. As such, it is characterised
by the six-vector
\begin{eqnarray}
 \Phi_\alpha & = & \epsilon_{\alpha\beta\gamma\delta\mu\eta} \, E^a_\beta \, E^b_\gamma \, E^c_\delta \, E^d_\mu \, B_\eta \nonumber \\
 & = & \epsilon_{\alpha\beta\gamma\delta\mu\eta} \, \epsilon_{\eta\nu\lambda\rho\sigma\tau} \, E^a_\beta \, E^b_\gamma \, E^c_\delta \, E^d_\mu \, O_\nu \, E^a_\lambda \, E^b_\rho \, E^c_\sigma \, E^d_\tau \, ,
\label{optimal}
\end{eqnarray}
with an unknown normalisation factor.

\subsection{Iterative Geometric Approach}
\vspace*{-0.2cm}
The linear geometric approach described above and used in~\cite{Ellis:2010xm} entails choosing a
sample of points in the MSSM scenario of interest, fixing the phases to $0^\circ$ or $\pm180^\circ$ for each scan point.
Next one computes the optimal direction using the above geometric approach, and then one
chooses randomly sets of phases along this direction. This is suitable as long as the phases are small, but we are
also interested in the possibilities for large phases.

Here we use an iterative approach to extend and improve the efficiency of the linear geometric approach.
After fixing the phases to $0^\circ$ or $\pm180^\circ$ and computing the favoured direction with the geometric approach
as discussed above, we move by $20^\circ$ along the favoured direction, and then
recompute the favoured direction at this new position. This procedure is then iterated up to $100^\circ$.
\vspace*{-0.4cm}

\subsection{Sampling Strategy}
\vspace*{-0.2cm}
We have generated several million points in each of the MSSM scenarios studied in the next Section.
Among those points, we have retained only those for which one of the neutral Higgs bosons
has a mass in the range 121-129 GeV (corresponding to the measured value $\simeq 125$~GeV
with a generous theoretical uncertainty), and we require the LSP to be the lightest neutralino.
In addition, we impose the LEP and Tevatron SUSY mass limits and require squarks and the gluino to have
masses above 500 GeV as a conservative implementation of
the LHC SUSY limits. Although the LHC SUSY search limits are stronger in
more constrained MSSM scenarios, they become weaker in more general scenarios such as the 
pMSSM~\cite{Arbey:2011aa,Arbey:2013iza,Cahill-Rowley:2014twa}. For consistency, 
here we apply the same loose constraints on the squark and gluino masses in all studied scenarios.
Other constraints, such as those imposed by heavy-flavour, Higgs and direct dark matter
measurements, are imposed at later stages in the analyses.

The SUSY mass spectra and couplings, as well as the EDM constraints, are computed with {\tt CPsuperH}~\cite{Lee:2003nta,Lee:2007gn,Lee:2012wa}. The thorium monoxide EDM is calculated using the following formula \cite{Cheung:2014oaa}:
\begin{equation}
 d_{\rm ThO} [e.{\rm cm}] / \mathcal{F}_{\rm ThO} = d_e [e.{\rm cm}] + 1.6\times10^{-21}[e.{\rm cm}] \; C_S\;{\rm TeV}^2 + \cdots \, ,
 \label{eq:tholimit}
\end{equation}
where $d_e$ is the electron EDM and $C_S$ the coefficient of the CP-odd electron nucleon interaction, which is also present in the thallium EDM. The left hand side of Eq.~(\ref{eq:tholimit}) is the quantity on which experimental constraints are provided currently\cite{Cheung:2014oaa}. Flavour constraints are calculated with {\tt SuperIso}~\cite{Mahmoudi:2007vz,Mahmoudi:2008tp} and {\tt CPsuperH}. For the calculation of the dark matter relic density we used {\tt SuperIso Relic}~\cite{Arbey:2009gu} and {\tt micrOMEGAs}~\cite{Belanger:2006is,Belanger:2008sj,Belanger:2013oya}, and the later is also used for the calculation of scattering cross sections for dark matter direct detection. Finally, we use {\tt HiggsBounds}~\cite{Bechtle:2013wla} to assess the viability of the model points in view of the Higgs constraints.

\section{Studies of MSSM Scenarios}

We now apply the approach described above to several representative MSSM scenarios.

\subsection{The CMSSM}
\label{sec:cmssm-bf}

We first consider the CMSSM, in which the soft supersymmetry-breaking parameters $m_0, m_{1/2}$
and $A$ are each constrained to have universal values at an input grand-unification scale. This model
is often analysed assuming some fixed value of $\tan \beta$, the ratio of Higgs v.e.v.s. Generalising the usual CMSSM set-up, here we vary the six MSSM CP phases independently in order to allow more flexibility
and a closer comparison with the other MSSM scenarios. Our starting-point here is one of the best-fit CMSSM points
found recently in a global analysis~\cite{Buchmueller:2013rsa} of the
$m_0, m_{1/2}, A, \tan \beta$ parameter space for the Higgsino mixing parameter $\mu > 0$,
neglecting all the possible MSSM sources of CP violation~\footnote{This analysis also found a
high-mass best-fit point with a slightly lower value of the global $\chi^2$ function. However, the
ATLAS jets + missing transverse energy constraint has subsequently been revised, and the low-mass point is
now the global minimum of the $\chi^2$ function~\cite{Buchmueller:2014yva}.}. This point has
\begin{equation}
m_0 \; = \; 670~{\rm GeV}, \; m_{1/2} \; = \; 1040~{\rm GeV}, \; A \; = \; 3440~{\rm GeV}, \; \tan \beta \; = \; 21 \, .
\label{CMSSMBF}
\end{equation}
We use this point as a base for the geometric approach using the EDM limits in Table~\ref{tab:EDMs},
treating the CP asymmetry in $b \to s \gamma$, $A_{CP}$, as the observable to be maximised,
in a follow-up of the study presented in \cite{Brooijmans:2014eja}.
We have generated more than 600000 sets of phases along the favoured direction, and have found that
about half of them pass the EDM constraints, which shows that the method is very efficient. 

\begin{figure}[!p]
 \begin{center}
  \includegraphics[width=7.5cm]{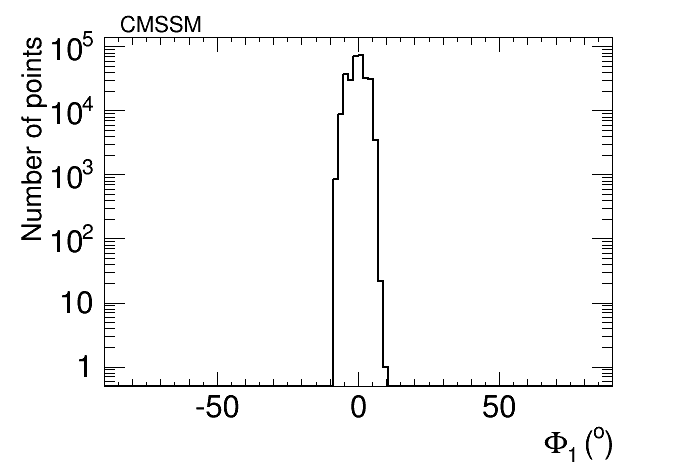}
  \includegraphics[width=7.5cm]{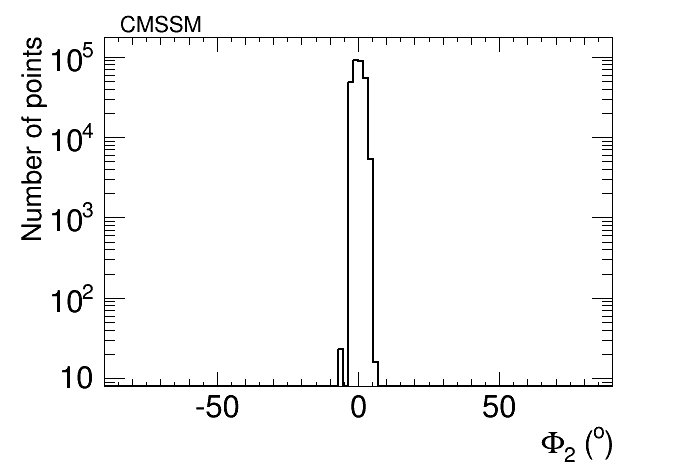}\\
  \includegraphics[width=7.5cm]{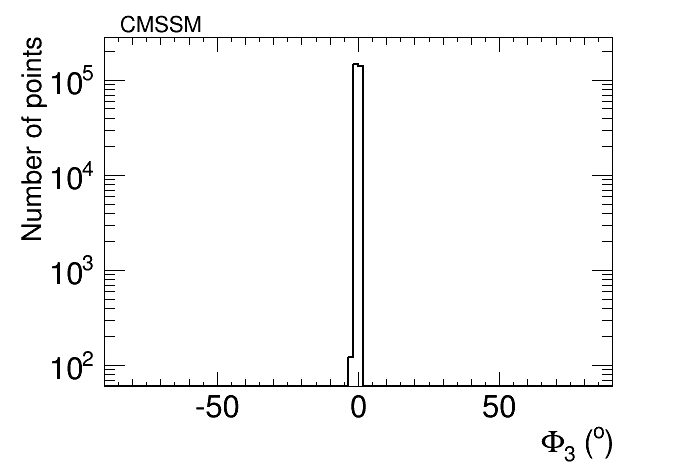}
  \includegraphics[width=7.5cm]{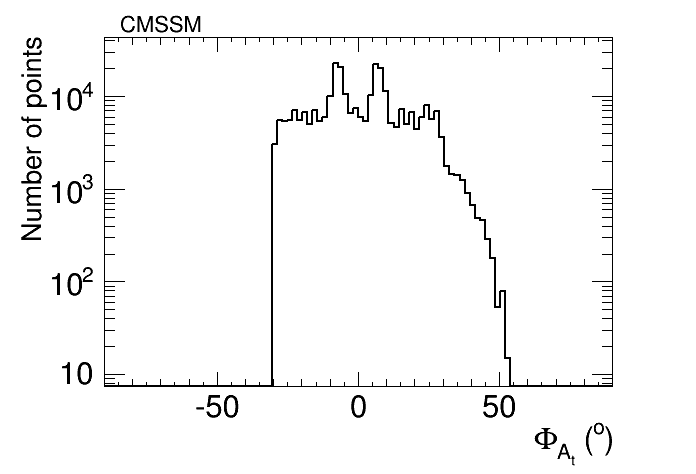}\\
  \includegraphics[width=7.5cm]{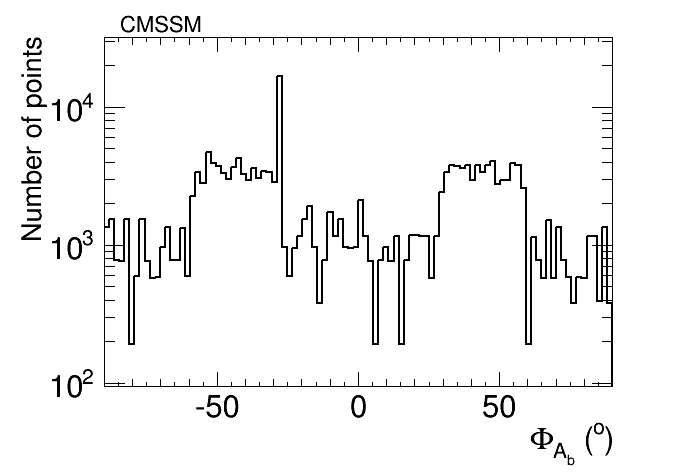}
  \includegraphics[width=7.5cm]{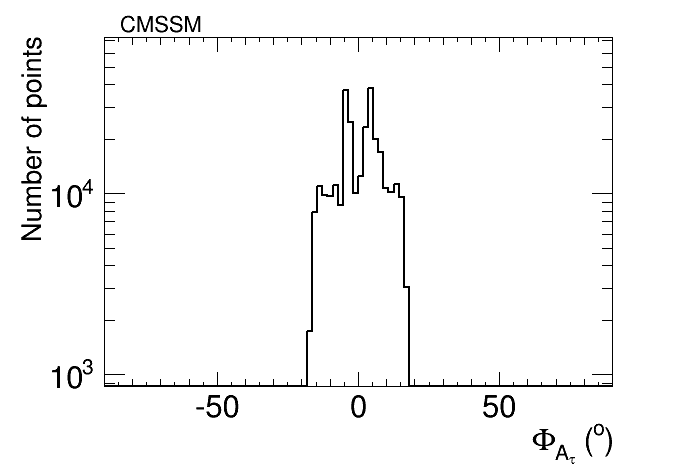}
\caption{\it Sampling of the CP-violating phases $\Phi_\alpha$  in the best-fit CMSSM scenario
(\protect\ref{CMSSMBF}) generated in the iterative
geometric approach, imposing the EDM limits as well as constraints from the
cosmological cold dark matter relic density and upper limits on direct detection,
from flavour physics and from Higgs searches.\label{fig:CMSSM-distrib}}
 \end{center}
\end{figure}

Fig.~\ref{fig:CMSSM-distrib} shows the distributions of the six CP-violating phases $\Phi_\alpha$
obtained from our sampling. The reader should bear in mind that
these distributions have no `probability' or `likelihood' interpretation,
but only indicate how our iterative geometric procedure samples large values of the phases. We see that the
effectiveness of the procedure differs significantly for different phases. 
We see that large values of $\Phi_{A_b}$ are relatively well sampled, 
whereas only intermediate $\Phi_{A_t}$ and $\Phi_{A_\tau}$ values can be reached, 
and we find no parameter sets with $\Phi_{1,2,3}$ substantially different from zero.
This is because for the CMSSM best-fit point (\ref{CMSSMBF}) it is not
possible to cancel the contributions of the phases to all the EDMs simultaneously.

\begin{figure}[!t]
 \begin{center}
\includegraphics[width=7.5cm]{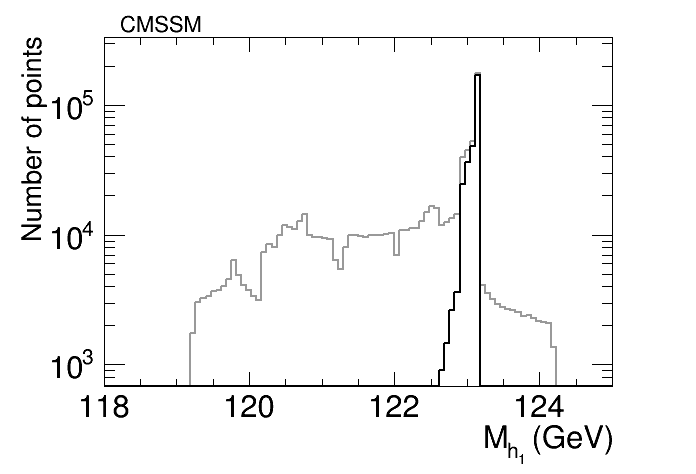}
\includegraphics[width=7.5cm]{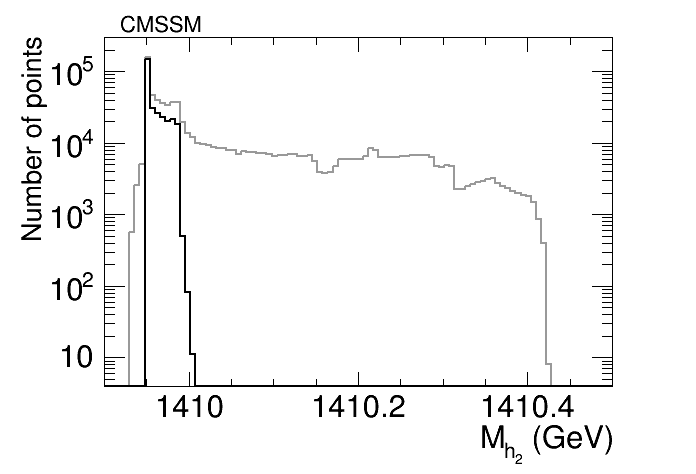}\\
\includegraphics[width=7.5cm]{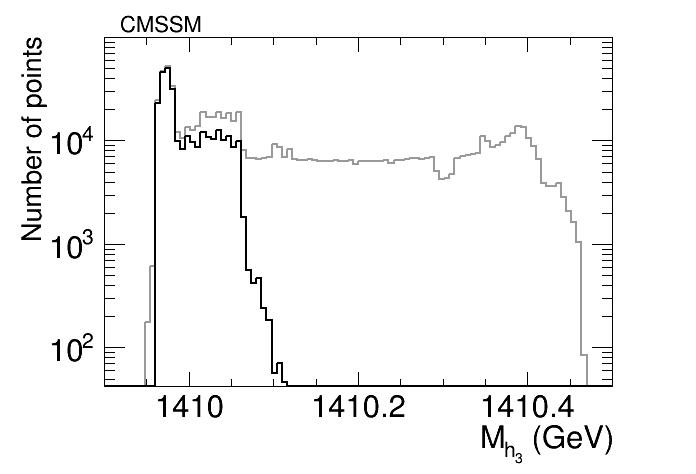}
\caption{\it The distributions of the three Higgs masses $M_{h_1},M_{h_2},M_{h_3}$ in the
CP-violating CMSSM before (grey) and after (black) applying the EDM constraints using the geometric approach
described in the text, assuming the best-fit values of $m_0, m_{1/2}, A$ and $\tan \beta$
(\protect\ref{CMSSMBF}) found in a global analysis.\label{fig:mh}}
 \end{center}
\end{figure}

Fig.~\ref{fig:mh} displays the results of this scan of the CP-violating CMSSM for the masses of the
three neutral Higgs bosons $M_{h_1},M_{h_2},M_{h_3}$. The Higgs masses all lie in
narrow ranges around their nominal values at the best-fit point in the CP-conserving CMSSM,
namely $M_h = 123$~GeV, $M_A \simeq M_H = 1410$~GeV. In view of the theoretical uncertainties
in calculating the Higgs masses for any specific set of CMSSM inputs, measuring Higgs masses
would not constrain usefully the CP-violating parameters at the CMSSM best-fit point.

\begin{figure}[!t]
 \begin{center}
 \includegraphics[width=7.5cm]{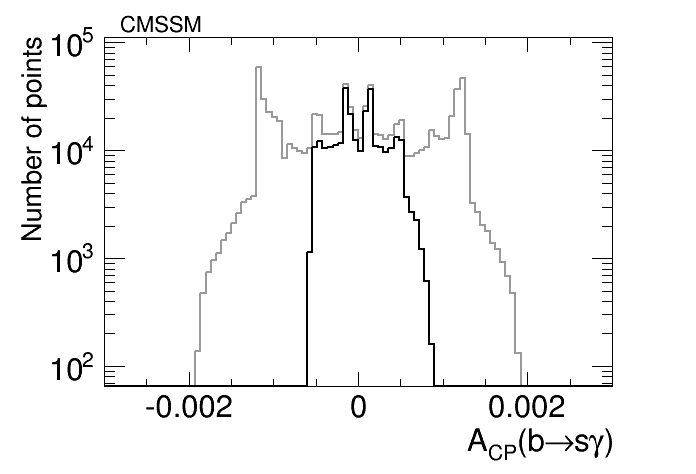}
 \includegraphics[width=7.5cm]{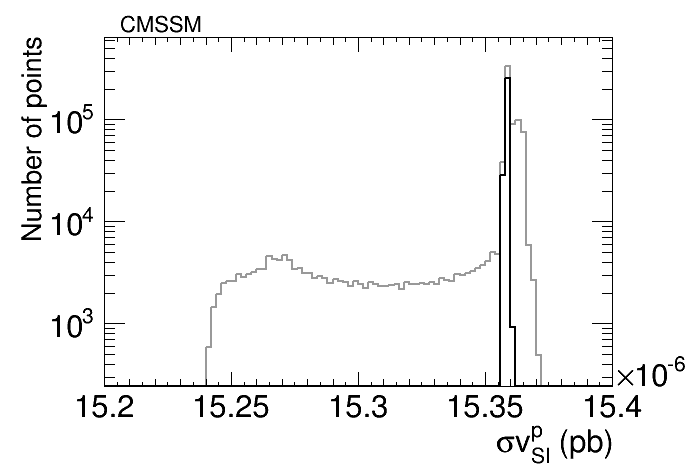}

\caption{\it The distributions of (left) the CP asymmetry in $b\to s\gamma$, $A_{CP}$, in the CMSSM
and (right) the spin-independent scattering cross-section for neutralino scattering
on protons, found before (grey) and after (black) applying the EDM constraints using the geometric approach
described in the text and assuming the best-fit values of $m_0, m_{1/2}, A$ and $\tan \beta$
(\protect\ref{CMSSMBF}) found in a global 
analysis~\protect\cite{Buchmueller:2013rsa,Buchmueller:2014yva}.\label{fig:ddpSI_ACPbsg}}
 \end{center}
\end{figure}

Fig.~\ref{fig:ddpSI_ACPbsg} displays the results of this CMSSM scan for the
CP asymmetry in $b\to s\gamma$, $A_{CP}$, (left) and the spin-independent 
neutralino-proton scattering cross-section $\sigma_{SI}^p v$ (right). We find in this model
values of $A_{CP} \ll 10^{-3}$, which are considerably below the current and prospective
experimental sensitivities. We conclude that the prospects for discovering the $A_{CP}$
signature of CP violation in this particular CMSSM scenario are not good.
Also, the spread in the values of $\sigma_{SI}^p v$ is quite small, and much smaller than
the theoretical uncertainties related to hadronic matrix elements and the astrophysical
uncertainties in the local dark matter density, so this observable is also not a promising one
for the CP-violating CMSSM.

We have also studied the possibility of a signature in $B_s$ meson mass mixing, with discouraging
results. We have found that the new physics contribution, $\Delta M^{NP}_{B_s}$ is always
very small, namely $\sim 0.1$/ps, which is far below any prospective reduction in the uncertainty
in the theoretical calculation of the contribution from the Standard Model~\cite{Lenz:2011ti}. Moreover, after applying
the EDM constraints the CP-violating CMSSM contribution is forced to be exceedingly close
to the value in the CP-conserving CMSSM.

\begin{figure}[!t]
 \begin{center}
  \includegraphics[width=7.5cm]{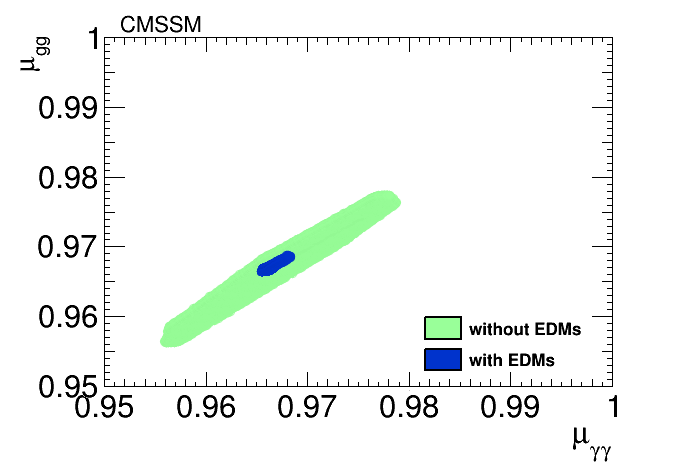}
  \includegraphics[width=7.5cm]{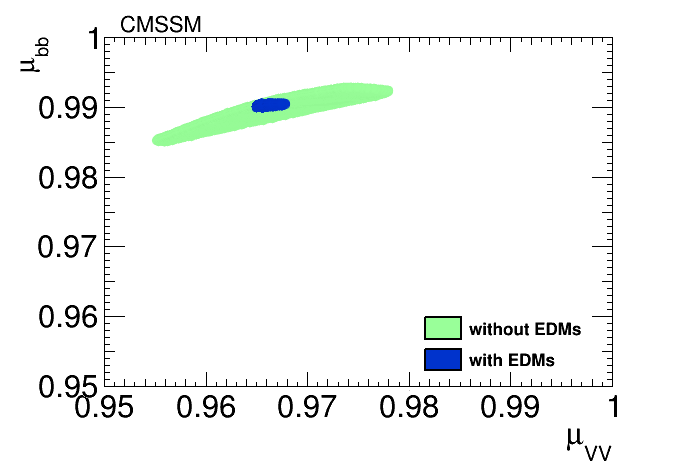}

\caption{\it Scatter plots of the $h_1$  signal strengths in the best-fit CMSSM scenario (\protect\ref{CMSSMBF})
in the CP-violating limit before (green dots) and after (blue dots) imposing the EDM constraints.
The left panel displays a strong linear correlation between $\mu_{\gamma \gamma}$
and $\mu_{gg}$, and the right panel displays the correlation between $\mu_{VV}$ and $\mu_{{\bar b}b}$.
\label{fig:CMSSM-muhiggs}}
 \end{center}
\end{figure}

In Fig.~\ref{fig:CMSSM-muhiggs} we show scatter plots of $h_1$ signal strengths $\mu_X$
(normalised relative to the Standard Model values) in the best-fit
CMSSM scenario (\ref{CMSSMBF}) with non-zero CP-violating phases before (green dots) and after (blue dots) the EDM constraints.
We see that the CP-violating case expands the ranges of these observables found
already in the CP-conserving case, in particular after imposing the EDM constraints. However, these
expanded ranges all lie well within the current experimental uncertainties.
In the left panel we see a strong, almost linear correlation between $\mu_{\gamma \gamma}$ and $\mu_{gg}$,
which becomes milder in the right panel,
between $\mu_{VV}$ and $\mu_{{\bar b}b}$. The signal strengths are close to, but smaller than unity.

We emphasise that the Higgs  couplings to fermions provide the most unambiguous probe of its CP properties,
even more so when there is CP mixing, as the Higgs may couple to the CP-even and CP-odd fermion
states in a democratic manner.  In cases of the $h_{i}VV$ couplings, there are two effects of CP mixing in the Higgs sector.
One  is a reduction in the strength of coefficient of the $g_{\mu \nu}$ term in the $h_{i} VV$ vertex and thus in the rates. 
The second is the simultaneous presence of the CP-even and CP-odd tensor structures in the vertex.
The coefficient of the CP-odd term in the $h_{i} VV$ vertex involving the $\epsilon_{\mu \nu \rho \lambda}$
tensor is by necessity small, as it is always loop-induced. Reduction in the production rates is reflected in signal strengths,
but these, while currently providing the best available information, are necessarily ambiguous,
as there are other mechanisms that may lead to the rate modification. On the other hand, since the fermions couple democratically to the  CP-even and CP-odd parts
of the Higgs couplings, ascertaining the simultaneous presence of $\bar f f h_{i}$ and $\bar f \gamma_5  f h_{i}$ terms
in the vertex through various angular distributions and kinematic variables is unambiguous. 

\begin{figure}[!t]
 \begin{center}
 \includegraphics[width=7.5cm]{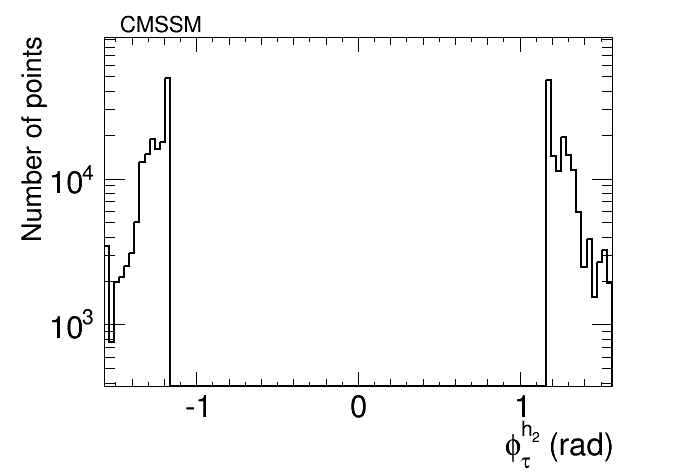}\includegraphics[width=7.5cm]{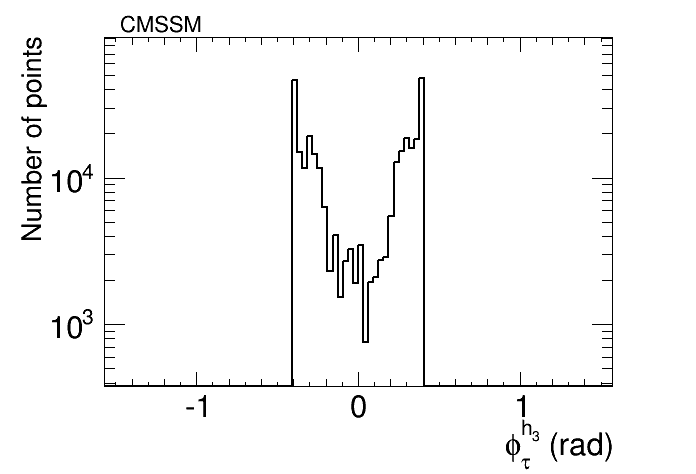}

\caption{\it The distributions of (left) the CP-violating phase $\phi_\tau^{h_2}$ in $h_2 \tau \tau$
couplings and (right) the CP-violating phase $\phi_\tau^{h_3}$ in $h_3 \tau \tau$
couplings in the CMSSM, found after  applying the EDM
constraints using the geometric approach
described in the text and assuming the best-fit values of $m_0, m_{1/2}, A$ and $\tan \beta$
(\protect\ref{CMSSMBF}) found in a global 
analysis~\protect\cite{Buchmueller:2013rsa,Buchmueller:2014yva}.\label{fig:CMSSMtau}}
 \end{center}
\end{figure}

\begin{figure}[!t]
 \begin{center}
 \includegraphics[width=7.5cm]{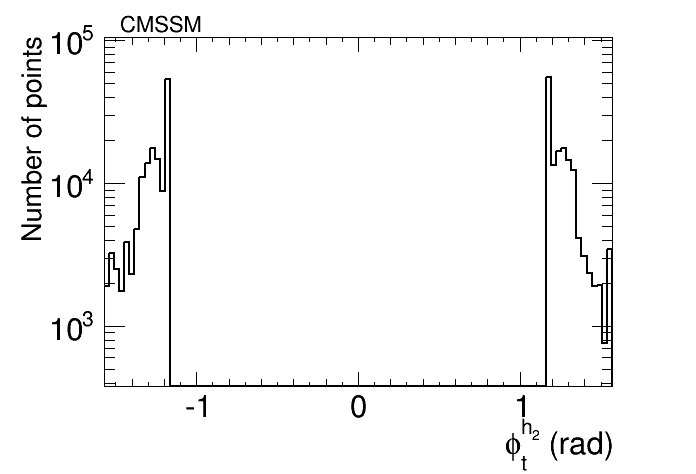}\includegraphics[width=7.5cm]{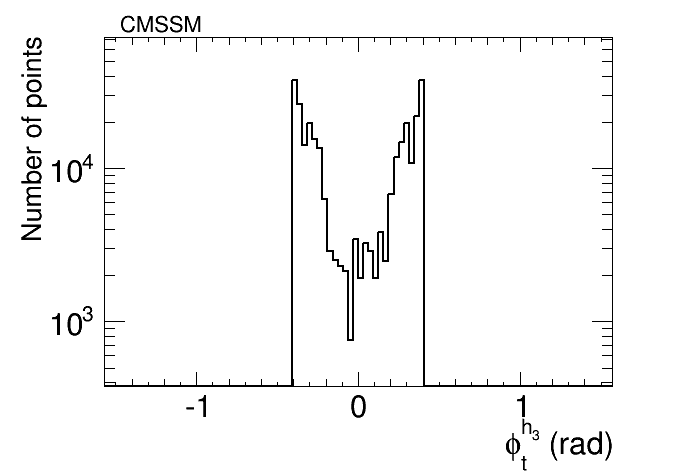}
\caption{\it The distributions of (left) the CP-violating phase $\phi_t^{h_2}$ in $h_2 {\bar t} t$
couplings and (right) the CP-violating phase $\phi_t^{h_3}$ in $h_3 {\bar t} t$
couplings in the CMSSM, found after  applying the EDM
constraints using the geometric approach
described in the text and assuming the best-fit values of $m_0, m_{1/2}, A$ and $\tan \beta$
(\protect\ref{CMSSMBF}) found in a global 
analysis~\protect\cite{Buchmueller:2013rsa,Buchmueller:2014yva}.\label{fig:CMSSMt}}
 \end{center}
\end{figure}

We have therefore analysed the prospects for CP violation in the couplings of the neutral Higgs
bosons to $\tau^+ \tau^-$ and ${\bar t} t$, by calculating the quantities $\phi^{h_i}_\tau$ and $\phi^{h_i}_t$
for $i = 1, 2, 3$, which are expressed in terms of the corresponding pseudoscalar and scalar couplings by
\begin{equation}
\tan \phi^{h_i}_\tau \; \equiv \; \frac{g_P^{h_i \tau \tau}}{g_S^{h_i \tau \tau}}\;, \quad
\tan \phi^{h_i}_t \; \equiv \; \frac{g_P^{h_i {\bar t} t}}{g_S^{h_i {\bar t} t}} \; .
\label{CPVheavyH}
\end{equation}
After imposing the EDM constraints, we find that the phases for the $h_1$ couplings are very small,
$\la 0.02$~radians. On the other hand, the phases for the $h_2$ and $h_3$ couplings may be
quite large, as seen in Figs.~\ref{fig:CMSSMtau} and \ref{fig:CMSSMt}, respectively.
The $h_2$ couplings have phases close to $\pm \pi$, corresponding to a mainly CP-odd state, while the $h_3$ couplings are close to 0 corresponding to a mainly CP-even state.
A detailed discussion of the prospects for measuring these phases at the LHC and/or
future colliders lies beyond the scope of this work. Clearly, any such future analysis would need
to take into account the near-degeneracy of the $h_2$ and $h_3$ bosons, as seen in
Fig.~\ref{fig:mh}, whose implications would be different for $pp$, $e^+ e^-$ , $\mu^+ \mu^-$ and $\gamma \gamma$
colliders. 

We limit ourselves to pointing out a few of these.
In the case of the light Higgs, associated production of Higgs with a ${\bar t} t$ pair or a single $t$
or ${\bar t}$ can be used for this~\cite{lighthcpv}.
However, in our case since it is the heavier Higgses that have the larger CP violation,
associated production may not be the best way, but  decays of the Higgs into a $\tau$ pair (or even into a ${\bar t} t$ pair if the
Higgs is  heavy enough) and analysis of the spins of the decay $\tau/t$  can be used at
$\gamma \gamma$ colliders\cite{gagacpv} and even at the LHC~\cite{Ellis:2004fs,Berge:2011ij,Chakraborty:2013si}.
The method of Ref.~\cite{Ellis:2004fs} is particularly promising when the $h_{2,3}$ are degenerate.

\subsection{NUHM2}
\label{sec:nuhm}

We now use the iterative geometric approach with four EDM constraints to
analyse CP violation in the NUHM2 scenario, in which the gaugino masses,
trilinear couplings and soft supersymmetry-breaking
contributions to the squark and slepton masses $M_{\tilde f}$ are universal, but those to the two
Higgs doublets are allowed to vary independently. The freedom in these two parameters can be
traded via the electroweak vacuum conditions for free values of $\mu$ and a heavy Higgs mass
parameter: to avoid complications with the three-way CP-violating mixing in the neutral sector, we
take this second free parameter to be $m_{H^\pm}$. We perform a random scan
over the following ranges of the NUHM2 mass parameters:
\begin{eqnarray}
M_1 \; = \; M_2 \; = \; M_3 \; = \; m_{1/2} \in [50,3000]~{\rm GeV}, & A_0 \in [0,10000]~{\rm GeV}, & \nonumber \\
M_{\tilde f} \; = \; m_0 \in [50,3000]~{\rm GeV}, \; m_{H^\pm} \in [1,2000]~{\rm GeV}, & \mu \in [-2000,2000]~{\rm GeV}, &
\label{NUHM2ranges}
\end{eqnarray}
with $\tan\beta \in [1,60]$ and varying the six phases $\Phi_\alpha$ independently as before, using the geometrical approach
to seek maximal values of $A_{CP}$.

\begin{figure}[!p]
 \begin{center}
  \includegraphics[width=7.5cm]{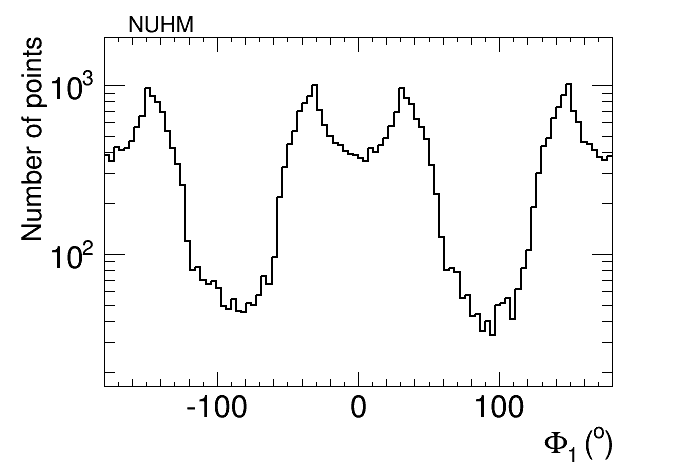}\includegraphics[width=7.5cm]{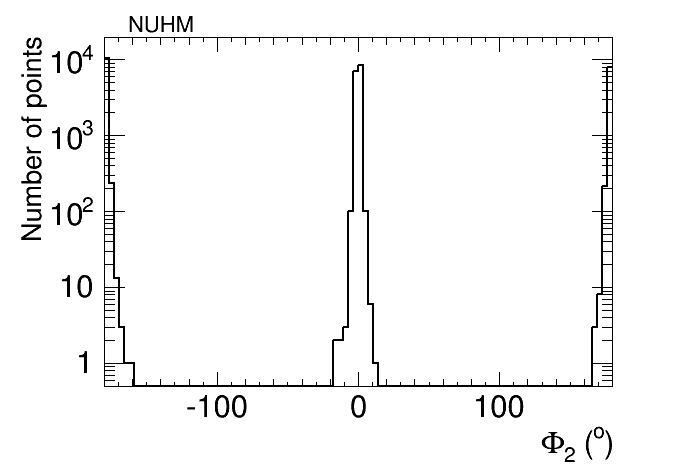}\\
  \includegraphics[width=7.5cm]{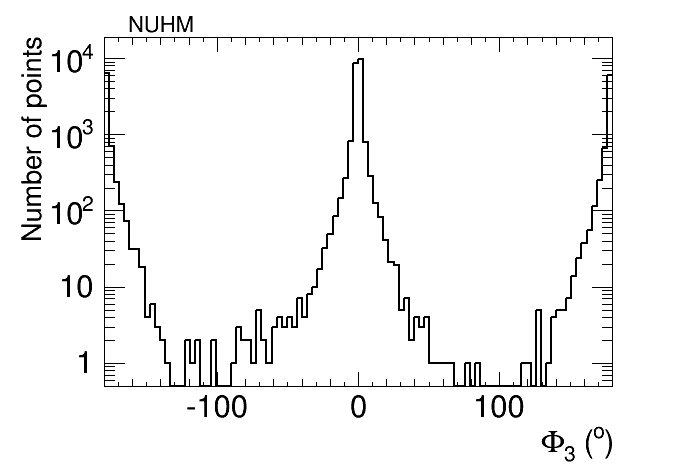}\includegraphics[width=7.5cm]{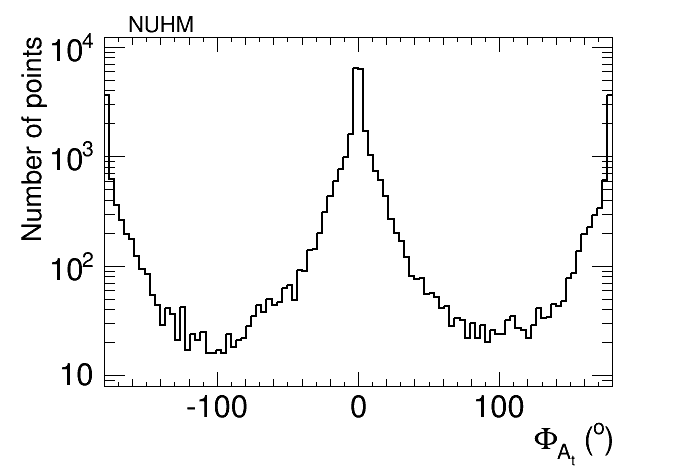}\\
  \includegraphics[width=7.5cm]{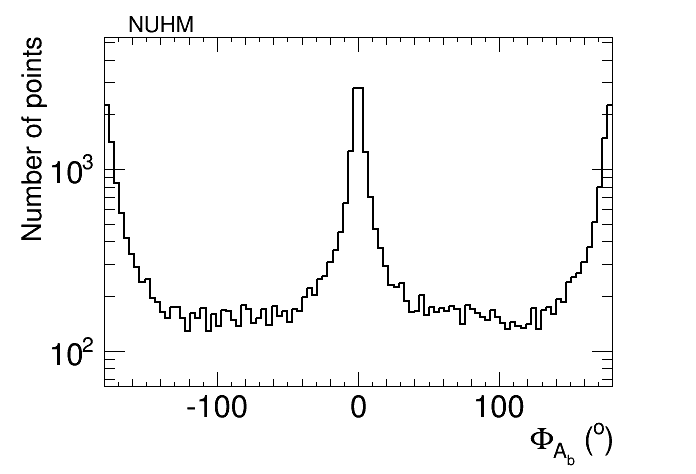}\includegraphics[width=7.5cm]{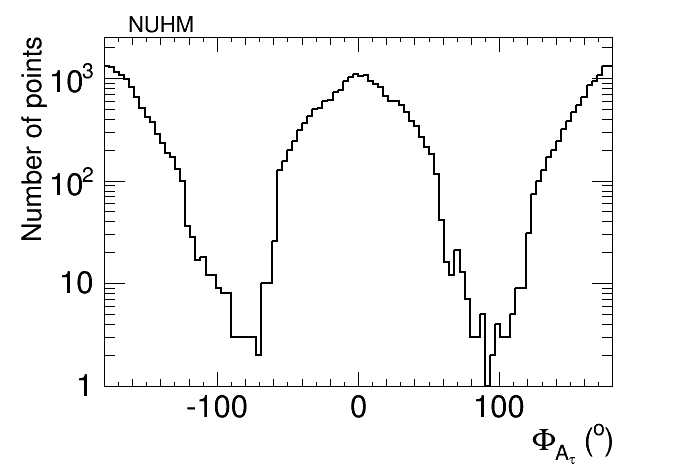}
\caption{\it Sampling of the CP-violating phases $\Phi_\alpha$  in the NUHM2 scenario generated in the iterative
geometric approach, imposing the EDM and other constraints.\label{fig:nuhm-distrib}}
 \end{center}
\end{figure}

Fig.~\ref{fig:nuhm-distrib} displays the samples of the six CP-violating phases $\Phi_\alpha$ obtained in our analysis.
We see that our iterative geometrical approach enables us to sample effectively large values of
$\Phi_1, \Phi_{A_t}$ and $\Phi_{A_b}$, whereas large values of $\Phi_3$ and $\Phi_{A_\tau}$
are sampled less effectively, and we do not find large values of $\Phi_2$. Again,
we emphasise that these distributions do not have any `probability' or `likelihood' interpretation.
However, the absence of large values of $\Phi_2$ indicates that there is no way
to cancel the contributions of this and the other phases to all the EDMs simultaneously.

\begin{figure}[!t]
 \begin{center}
\includegraphics[width=7.5cm]{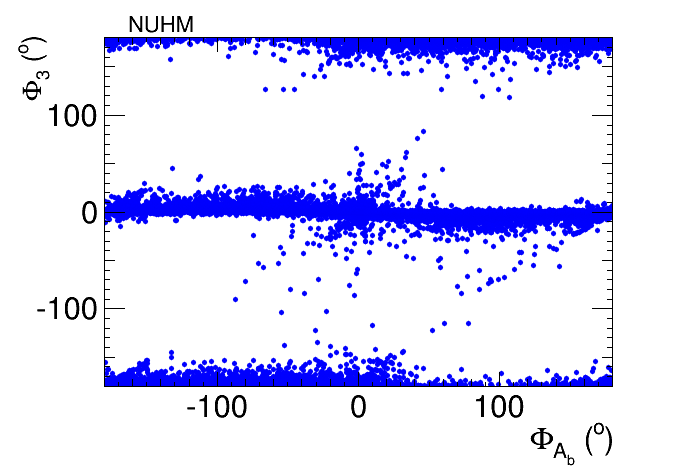}\includegraphics[width=7.5cm]{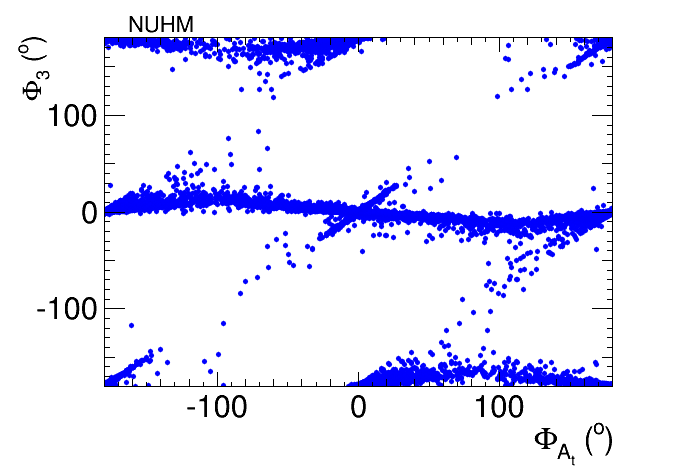}
\caption{\it Correlations of $\Phi_3$ with $\Phi_{\rm{A_b}}$ (left panel) and
$\Phi_3$ with $\Phi_{\rm{A_t}}$ (right panel) imposed by the EDM constraints in the NUHM2 scenario.}
\label{fig:NUHM2cancel}
 \end{center}
\end{figure}

Fig.~\ref{fig:NUHM2cancel} provides a visualisation of the cancellations that are required to respect
the EDM constraints. In the left panel we see the correlation these constraints impose between
$\Phi_3$ and $\Phi_{A_t}$, and in the right panel the correlation between $\Phi_3$ and $\Phi_{A_b}$.
In both cases we see diagonal features corresponding to close correlations, but we also see
populations of points with large phases, e.g., in the neighbourhood of $(\Phi_{A_t}, \Phi_3) \sim (90^\circ, -90^\circ)$
in the left panel, and extending to $(\Phi_{A_b}, \Phi_3) \sim (-90^\circ, -90^\circ)$ in the right panel.
These examples serve as reminders that the EDM constraints do not require all the CP-violating
phases to be small simultaneously.

\begin{figure}[!t]
 \begin{center}
  \includegraphics[width=7.5cm]{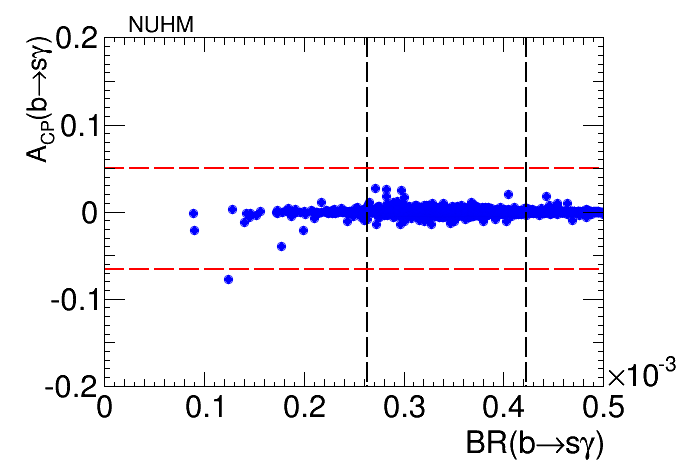} \includegraphics[width=7.5cm]{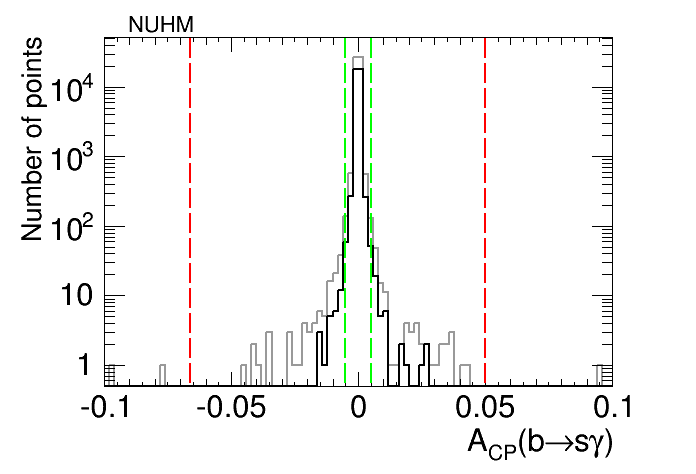}
\caption{\it Left panel: Scatter plot of the branching ratio for $b\to s\gamma$ decay versus its
CP-violating asymmetry, $A_{CP}$, in the NUHM2 scenario. Right panel: Histogram
of $A_{CP}$ in the NUHM2, imposing only the Higgs mass and EDM cuts (grey: full sample, black:
points satisfying the EDM constraints). The vertical red
dashed lines represent the present experimental limits, and the vertical green dashed lines
represent the prospective future improvement in the sensitivity to $A_{CP}$ by a factor of 10.\label{fig:nuhm-ACPbsg}}
 \end{center}
\end{figure}

The iterative geometric approach was designed to find the maximal values of the CP-violating
asymmetry in $b \to s \gamma$ decays, $A_{CP}$, that are compatible with the EDM constraints.
We see in the left panel of Fig.~\ref{fig:nuhm-ACPbsg} that values of $A_{CP} \la 2$\% can be found in the NUHM2 for values
of the $b \to s \gamma$ branching ratio lying within the experimentally allowed range. The right panel of
Fig.~\ref{fig:nuhm-ACPbsg} displays a histogram of these results for the NUHM2 (grey: full sample, black:
points satisfying the EDM constraints). The present experimental constraints
on $A_{CP}$ are shown as vertical red dashed lines~\cite{Agashe:2014kda},
and the vertical green dashed lines represent the
possible future improvement in the experimental sensitivity by a factor of 10,
corresponding to the prospective Belle II sensitivity~\cite{Aushev:2010bq}. We see that there are CP-violating
NUHM2 models that could be explored with such an improvement: the EDM constraints do not exclude an
observable value of $A_{CP}$, and such a measurement would provide additional information on CP violation
within the NUHM2.

\begin{figure}[!t]
 \begin{center}
  \includegraphics[width=8.5cm]{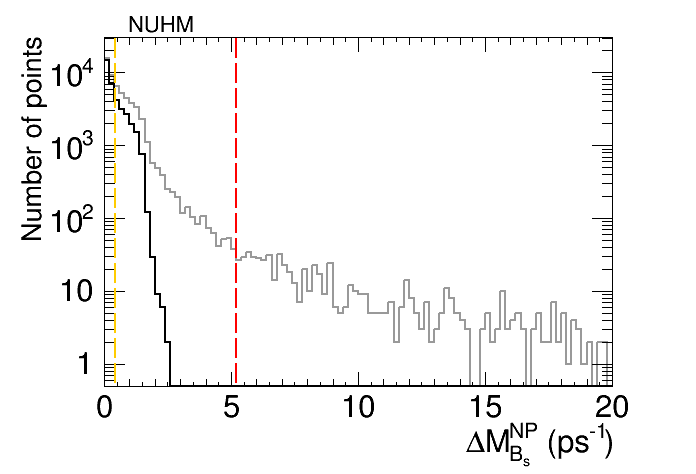}
\caption{\it Histogram of the possible new physics contribution to $B_s$ mixing, $\Delta M^{NP}_{B_s}$,
in the NUHM2 scenario. The grey histogram is for points satisfying the Higgs mass and other constraints,
and the black histogram is for points that also satisfy the EDM constraints. The vertical red dashed
line is the present experimental upper limit on $\Delta M^{NP}_{B_s}$, and the vertical yellow dashed line
shows the potential of a reduction in the current theoretical uncertainty in the Standard Model by a factor of 10.
\label{fig:NUHM2Bs}}
 \end{center}
\end{figure}

We have also calculated the possible new physics contribution to $B_s$ meson
mass mixing, $\Delta M^{NP}_{B_s}$, in the NUHM2 scenario, as shown in Fig.~\ref{fig:NUHM2Bs}. The grey
histogram is for the full sample of NUHM2 points satisfying the Higgs mass and other constraints, and the black histogram is for
points that also satisfy the EDM constraints. The present experimental upper limit on $\Delta M^{NP}_{B_s}$
is shown as the vertical red dashed line~\cite{Agashe:2014kda}.
The vertical yellow dashed line in Fig.~\ref{fig:NUHM2Bs}
represents the possible sensitivity if the theoretical uncertainty in the Standard Model contribution to $B_s$ mixing
could be reduced by a factor of 10 thanks to improved lattice calculations.
In this case, many of the viable NUHM2 models (indicated by the black histogram) could be explored.

\begin{figure}[!t]
 \begin{center}
  \includegraphics[width=7.5cm]{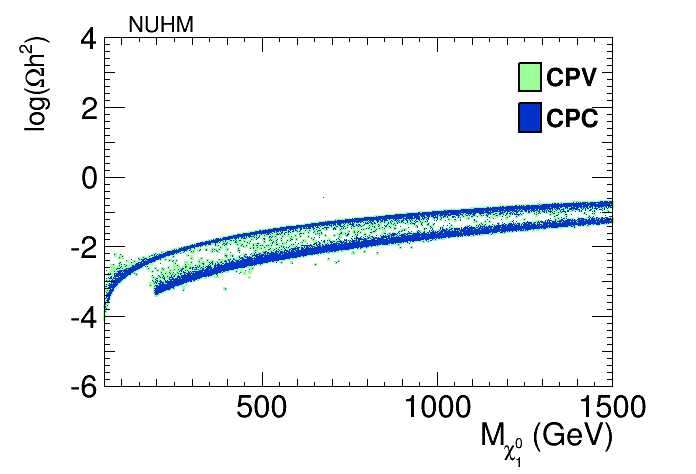}\includegraphics[width=7.5cm]{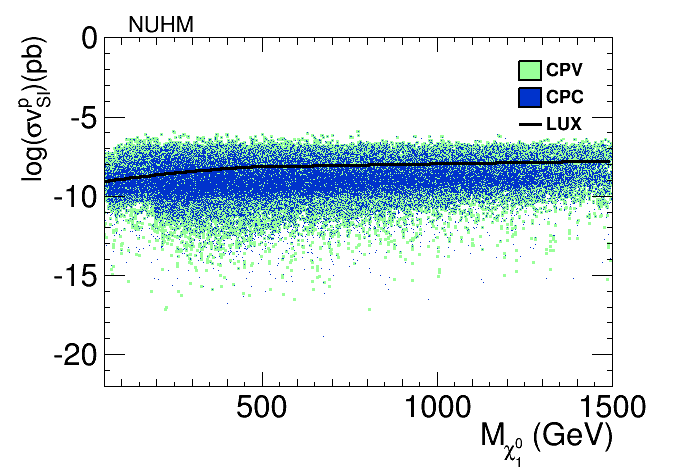}
\caption{\it Left panel: Scatter plot of the dark matter relic density as a function of the neutralino mass in the NUHM2 scenario.
Right panel: Scatter plot of the spin-independent dark matter scattering cross section
$\sigma_{SI}^p v$ as a function of the neutralino mass in the NUHM2 scenario.
In both panels, CP-conserving parameter choices are denoted by blue dots, and CP-violating parameter
choices by green dots. \label{fig:nuhm-DM}}
 \end{center}
\end{figure}

We have not imposed {\it a priori} consistency with the cosmological constraints on the
relic LSP density $\Omega_\chi h^2$ and the spin-independent dark matter scattering
cross section $\sigma_{SI}^p v$. As we see in the left panel of Fig.~\ref{fig:nuhm-DM},
the values of the relic density for the CP-violating NUHM2 (green points) are very similar
to those in the CP-conserving version (blue points), and they are generally within the
range allowed for a supersymmetric contribution to the dark matter density. The right
panel of Fig.~\ref{fig:nuhm-DM} shows that the values of $\sigma_{SI}^p v$ are also
rather similar, with some differences for low cross-section values well below the
experimental upper limit from LUX~\cite{Akerib:2013tjd}, which is shown as the black solid line.

\begin{figure}[!t]
 \begin{center}
  \includegraphics[width=7.5cm]{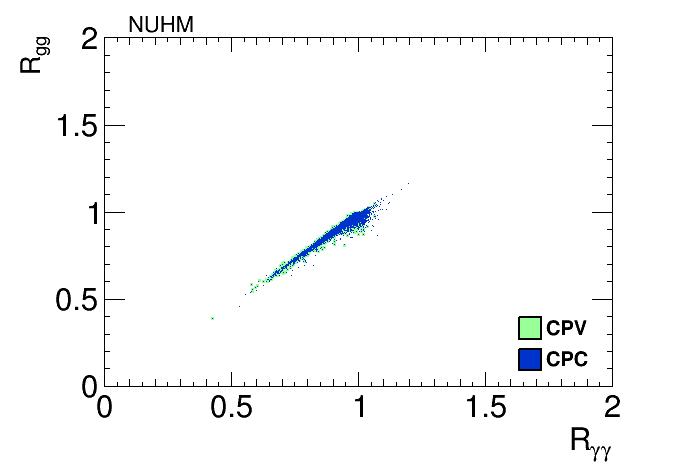}\includegraphics[width=7.5cm]{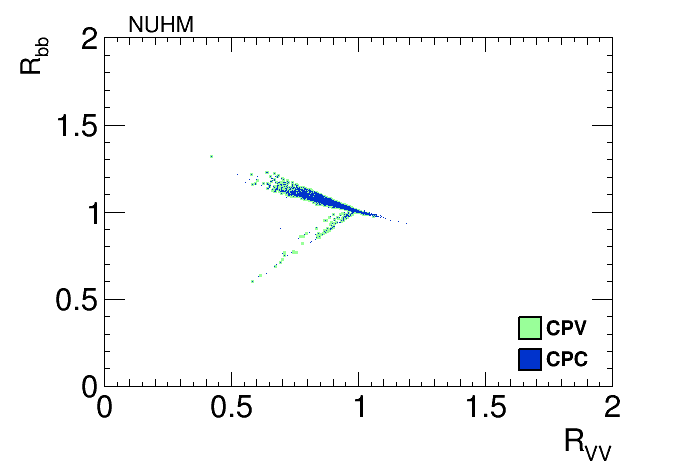}
\caption{\it Scatter plots of branching ratios, normalised to the Standard Model values,
for decays of the lightest Higgs boson, $h_1$,
in the NUHM2 scenario in the CP-violating limit $\Phi_\alpha = 0$ (blue dots) and in the CP-violating
sample (green dots). The left panel displays a linear correlation between $R_{\gamma \gamma}$
and $R_{gg}$, and the right panel displays a bimodal correlation between $R_{VV}$ and $R_{{\bar b}b}$.
\label{fig:nuhm-Rhiggs}}
 \end{center}
\end{figure}

Fig.~\ref{fig:nuhm-Rhiggs} shows scatter plots of values of $h_1$ branching ratios  in the NUHM2 scenario
The left panel displays $(R_{\gamma \gamma}, R_{gg})$ and the right panel displays $(R_{VV}, R_{{\bar b}b})$.
The blue dots are CP-conserving parameter choices with $\Phi_\alpha = 0$, and the green dots
are from a scan of CP-violating points with $\Phi_\alpha \ne 0$. We note in the left panel a strong correlation between
$R_{\gamma \gamma}$ and $R_{gg}$, which may be either much smaller than in the Standard Model or somewhat larger,
which is due to the variation of the Higgs width induced by a modification of the Higgs to ${\bar b}b$ 
branching fraction~\footnote{Small values of $R_{\gamma \gamma}$ and $R_{gg}$ are disfavoured by LHC Higgs measurements.}.
We see in the right panel that a large reduction in $R_{VV}$ is also possible,
which may be accompanied by values of $R_{{\bar b}b}$ that are either larger or smaller than
in the Standard Model. The branch with larger values of $R_{{\bar b}b}$
is also related to the variation of the Higgs width, while the points corresponding to a
decrease of both ratios are due to an enhancement of decays to light SUSY 
particles~\cite{AlbornozVasquez:2011aa,Arbey:2012dq,Arbey:2012bp}~\footnote{Small values of $R_{VV}$ are disfavoured by LHC Higgs measurements,
whereas a relatively large range of $R_{{\bar b}b}$ is still allowed.}.

\begin{figure}[!t]
 \begin{center}
  \includegraphics[width=7.5cm]{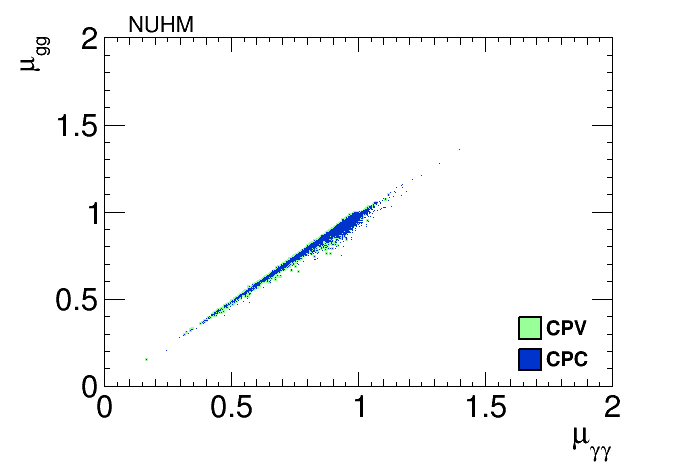}\includegraphics[width=7.5cm]{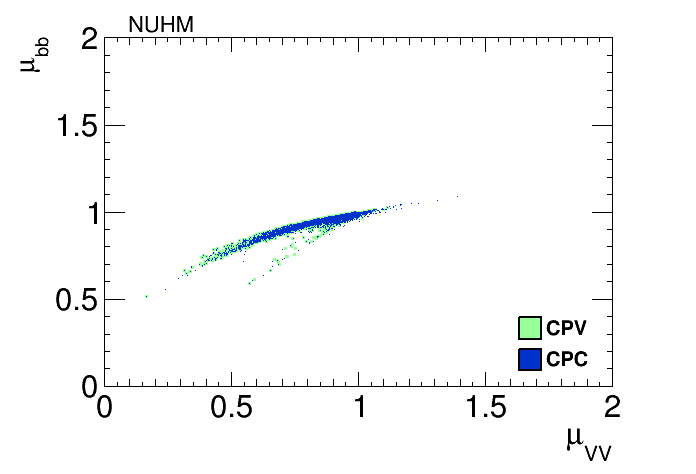}
\caption{\it Scatter plots of the $h_1$  signal strengths in the NUHM2 scenario in the CP-violating limit
$\Phi_\alpha = 0$ (blue dots) and in the CP-violating sample (green dots).
The left panel displays a strong linear correlation between $\mu_{\gamma \gamma}$
and $\mu_{gg}$, and the right panel displays a bimodal
correlation between $\mu_{VV}$ and $\mu_{{\bar b}b}$.\label{fig:nuhm-muhiggs}}
 \end{center}
\end{figure}

Scatter plots of $h_1$ signal strengths $\mu_X$ in the NUHM2
scenario with the CP-violating phases $\Phi_\alpha =0$ (blue dots) and $\ne 0$ (green dots)
are shown in Fig.~\ref{fig:nuhm-muhiggs}.

We see a strong, almost linear correlation between $\mu_{\gamma \gamma}$ and $\mu_{gg}$ in the left panel,
and in the right panel we see a correlation between $\mu_{VV}$ and $\mu_{{\bar b}b}$ that is bimodal
for small values of $\mu_{VV}$. No significant difference is observed between the CP-conserving and the CP-violating cases.

\begin{figure}[!t]
 \begin{center}
 \includegraphics[width=7.5cm]{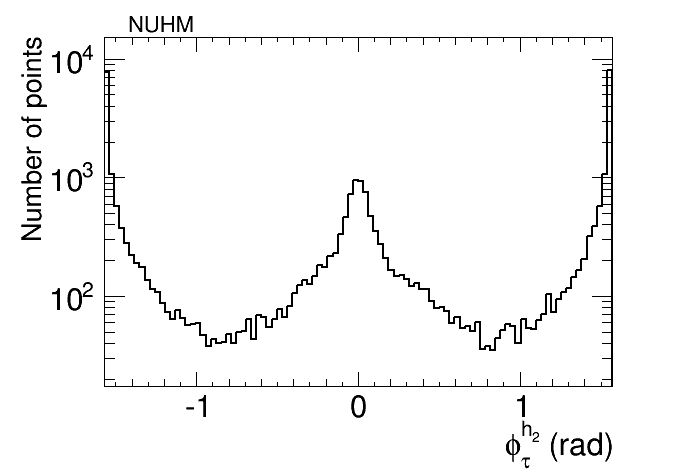}\includegraphics[width=7.5cm]{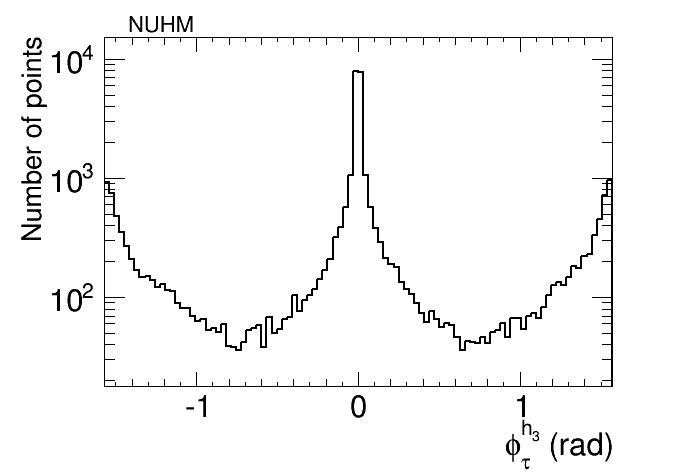}
\caption{\it The distributions of (left) the CP-violating phase $\phi_\tau^{h_2}$ in $h_2 \tau \tau$
couplings and (right) the CP-violating phase $\phi_\tau^{h_3}$ in $h_3 \tau \tau$
couplings in the NUHM2 scenario (\protect\ref{NUHM2ranges}),
found after  applying all the constraints using the geometric approach
described in the text.\label{fig:NUHM2tau}}
 \end{center}
\end{figure}

\begin{figure}[!t]
 \begin{center}
 \includegraphics[width=7.5cm]{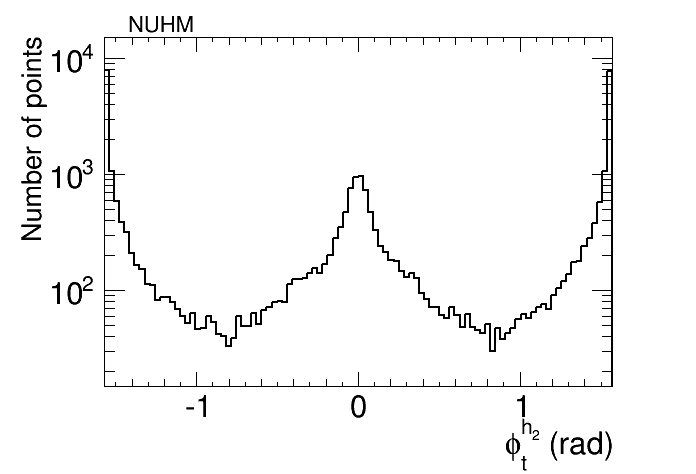}
 \includegraphics[width=7.5cm]{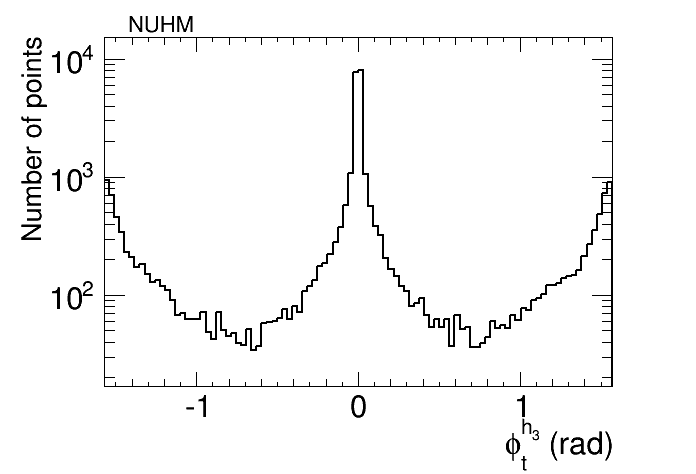}
\caption{\it The distributions of (left) the CP-violating phase $\phi_t^{h_2}$ in $h_2 {\bar t} t$
couplings and (right) the CP-violating phase $\phi_t^{h_3}$ in $h_3 {\bar t} t$
couplings in the NUHM2 scenario (\protect\ref{NUHM2ranges}),
found after  applying all the constraints using the geometric approach
described in the text.\label{fig:NUHM2t}}
 \end{center}
\end{figure}

The prospects for CP violation in the couplings of the heavy neutral Higgs
bosons to $\tau^+ \tau^-$ and ${\bar t} t$ in the NUHM2 scenario (\ref{NUHM2ranges})
are shown in Figs.~\ref{fig:NUHM2tau} and \ref{fig:NUHM2t}.
As in the CMSSM case discussed previously,
we find that after imposing all the constraints the phases for the $h_1$ couplings are small,
namely $\la 0.02$~radians.
On the other hand, $h_{2,3}$ decays may provide interesting
prospects for probing CP violation also in this NUHM2 scenario.

\subsection{CPX}
\label{sec:cpx}

We now apply the iterative geometric approach with four EDMs and one CP-violating observable described earlier
to a CPX scenario in which
\begin{eqnarray}
M_{Q_3} \; = \; M_{U_3} \; =\; M_{D_3} & = & M_{L_3} \; = \; M_{E_3} \; \equiv \; M_S \, , \nonumber \\
\mu \; = \; 4M_S\, , \; \; |A_{t,b,\tau} | \; = \; 2 M_S\, , & & |M_{1,2}| \; = \; 1~{\rm TeV}\, , \; \; |M_3| \; = \; 3~{\rm TeV} \, ,
\label{CPX}
\end{eqnarray}
performing random scans over the following parameter ranges:
\begin{equation}
M_S \; \in \; [50,3000]~{\rm GeV}\,, \; \;
m_{H^\pm} \; \in \; [1,2000]~{\rm GeV}\,, \;\;
\tan\beta \; \in \; [1,60] \, ,
\label{varyCPX}
\end{equation}
with the six CP-violating phases of the MCPMFV model being considered independent, as before.

\begin{figure}[!p]
 \begin{center}
  \includegraphics[width=7.5cm]{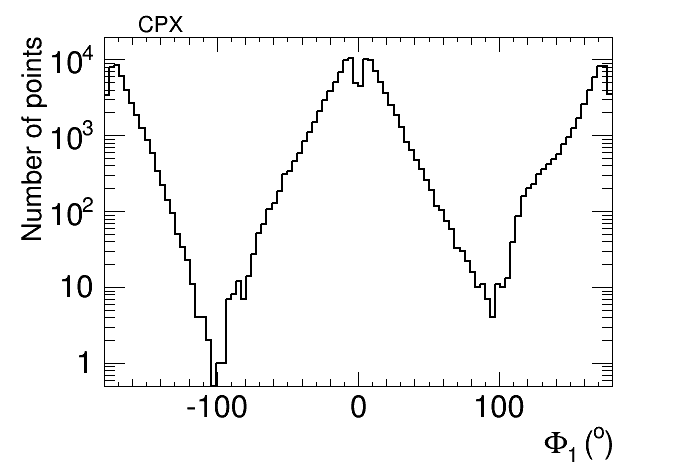}\includegraphics[width=7.5cm]{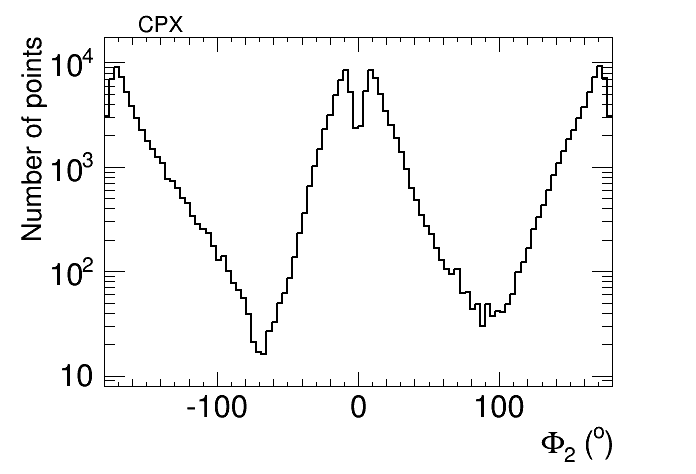}\\
  \includegraphics[width=7.5cm]{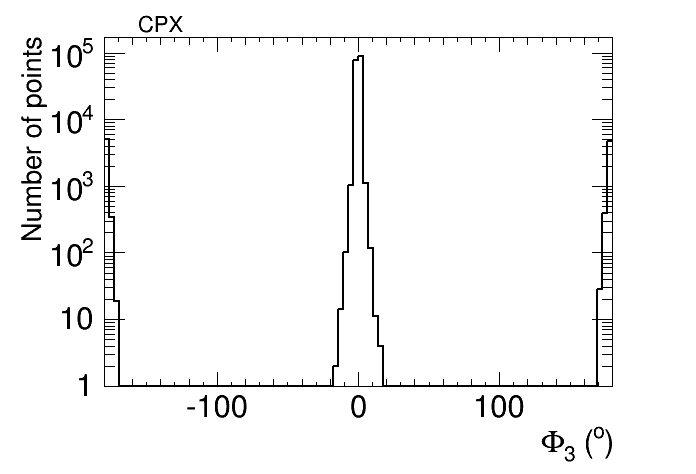}\includegraphics[width=7.5cm]{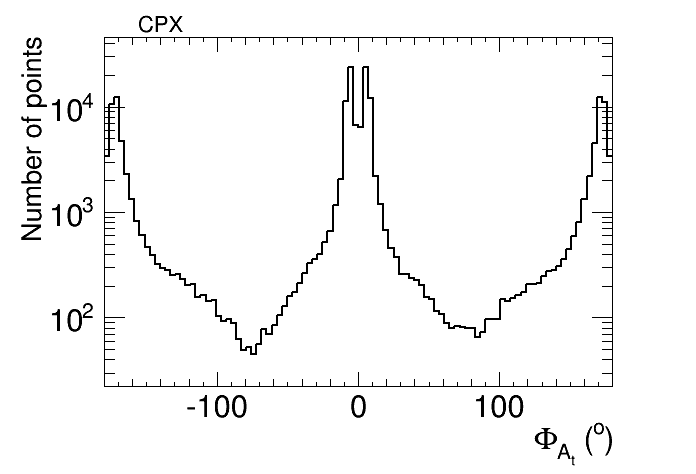}\\
  \includegraphics[width=7.5cm]{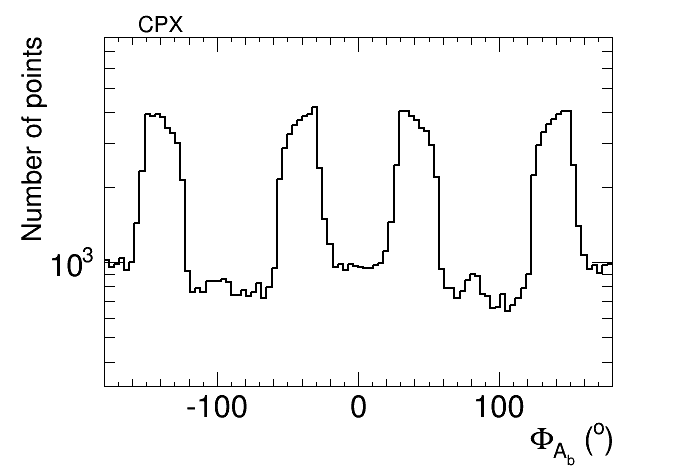}\includegraphics[width=7.5cm]{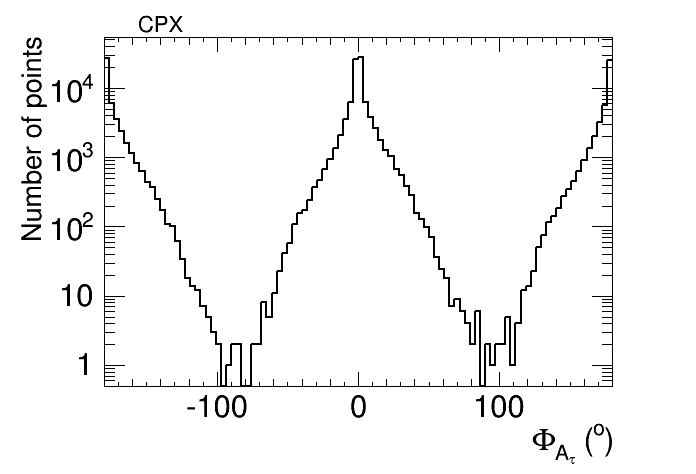}
\caption{\it Sampling of the CP-violating phases $\Phi_\alpha$  in the CPX scenario generated in the iterative
geometric approach, imposing the EDM and other constraints.\label{fig:cpx-distrib}}
 \end{center}
\end{figure}

Fig.~\ref{fig:cpx-distrib} displays the distributions of the six CP-violating phases $\Phi_\alpha$ sampled in our analysis.
We emphasise that these distributions do not have any `probability' or `likelihood' interpretation.
Rather, they serve to indicate how well our iterative geometric procedure gives access to large values of the phases
that are difficult to sample in a simple random scan, because of the cancellations required to
bring the EDMs within the allowed ranges shown in Table~\ref{tab:EDMs}. We see that the
effectiveness of the procedure differs significantly for different phases. For example, in the case of $\Phi_{A_b}$
our procedure yields almost as many parameter sets with $\Phi_{A_b} \sim \pm \ 90^\circ$ as with
$\Phi_{A_b} \sim 0^\circ$ or $180^\circ$, and actually yields {\it more} parameter sets with intermediate
values of $\Phi_{A_b}$. In the case of $\Phi_{A_t}$, the procedure yields a factor $\sim 100$ lower sampling
density for $\Phi_{A_b} \sim \pm \ 90^\circ$ than for $\Phi_{A_b} \sim 0^\circ, 180^\circ$, and larger factors
for $\Phi_2$, $\Phi_1$ and $\Phi_{A_\tau}$. Finally, we find no parameter sets for $\Phi_3 \sim \pm \ 90^\circ$:
this is because (for the choices of soft supersymmetry-breaking parameters in (\ref{CPX})) there is no way
to cancel the contributions of this and the other phases to all the EDMs simultaneously.

\begin{figure}[!t]
 \begin{center}
  \includegraphics[width=8.5cm]{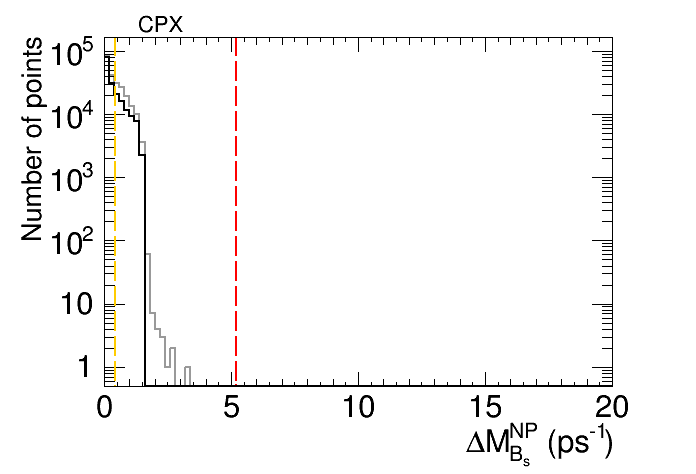}
\caption{\it Histogram of the possible new physics contribution to $B_s$ mixing, $\Delta M^{NP}_{B_s}$,
in the CPX scenario. The grey histogram is for points satisfying the Higgs mass and other constraints,
and the black histogram is for points that also satisfy the EDM constraints. The vertical red dashed
line is the present experimental upper limit on $\Delta M^{NP}_{B_s}$, and the vertical yellow dashed line
shows the potential of a reduction in the current theoretical uncertainty in the Standard Model by a factor of 10.
\label{fig:CPXBs}}
 \end{center}
\end{figure}

In the CPX scenario we do not find values of $A_{CP}$ that are large enough to be observable in the foreseeable
future. However, we do find a possible signature in the new physics contribution to $B_s$ meson
mass mixing, $\Delta M^{NP}_{B_s}$, as shown in Fig.~\ref{fig:CPXBs}. The grey
histogram is for CPX points satisfying the Higgs mass and other constraints, and the black histogram is for
points that also satisfy the EDM constraints, including the present experimental upper limit on $\Delta M^{NP}_{B_s}$,
which is shown as the vertical red dashed line. The magnitude of this upper limit is largely due to the theoretical
uncertainty in the Standard Model contribution to $B_s$ mixing, which is in turn associated with lattice calculations.
If this uncertainty could be reduced by a factor of 10, the sensitivity to new physics in $B_s$ mixing would
become that indicated by the vertical yellow dashed line in Fig.~\ref{fig:CPXBs}, which could explore many
of the CPX models indicated by the black histogram.

\begin{figure}[!t]
 \begin{center}
  \includegraphics[width=7.5cm]{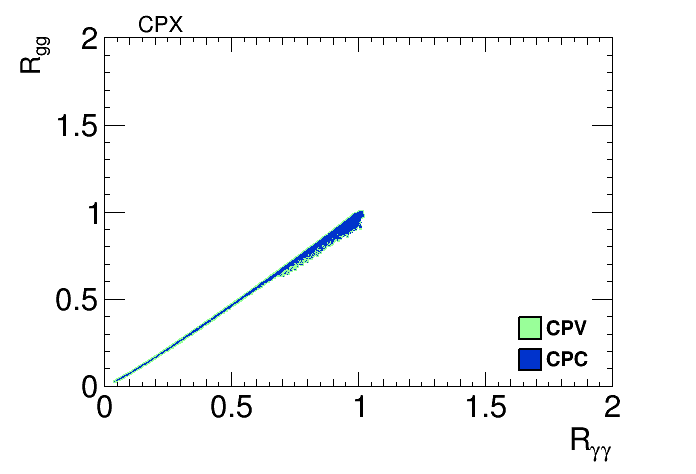}\includegraphics[width=7.5cm]{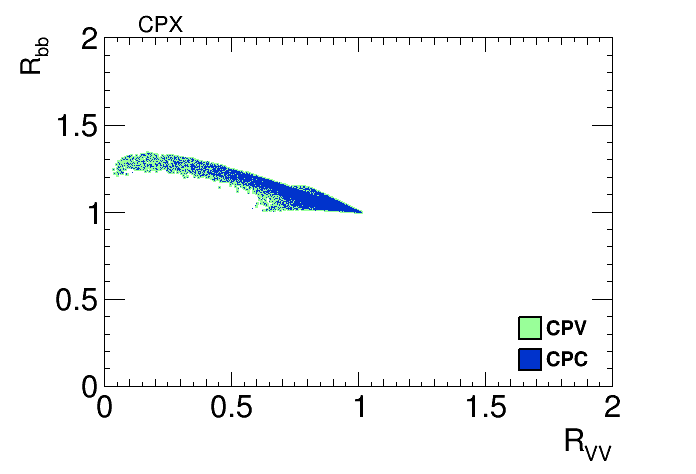}
\caption{\it Scatter plots of branching ratios, normalised to the Standard Model values,
for decays of the lightest Higgs boson, $h_1$,
in the CPX scenario in the CP-violating limit $\Phi_\alpha = 0$ (blue dots) and in the CP-violating
sample (green dots). The left panel displays a linear correlation between $R_{\gamma \gamma}$
and $R_{gg}$, and the right panel displays a non-linear anti-correlation between $R_{VV}$ and $R_{{\bar b}b}$.
\label{fig:cpx-Rhiggs}}
 \end{center}
\end{figure}

We display in Fig.~\ref{fig:cpx-Rhiggs} scatter plots of values of branching ratios  in the CPX scenario
of the lightest Higgs boson, $h_1$, normalised relative to the Standard Model values. The left panel shows
$(R_{\gamma \gamma}, R_{gg})$ and the right panel shows $(R_{VV}, R_{{\bar b}b})$
in the limits where the phases $\Phi_\alpha = 0$ (blue dots) and scanning
over the values of $\Phi_\alpha \ne 0$ allowed by the EDMs (green dots). There are
very small differences between the values of these quantities found in the CP-conserving
and CP-violating samples. In both cases, correlated substantial reductions in
$R_{\gamma \gamma}$ and $R_{gg}$ are possible, as is a large reduction in $R_{VV}$
relative to the Standard Model value. On the other hand, the `Cuba'-shaped plot in the right panel
shows that $R_{{\bar b}b}$ is anti-correlated with $R_{VV}$, and may be enhanced to $\sim 1.3$ times the Standard Model
value.

\begin{figure}[!t]
 \begin{center}
  \includegraphics[width=7.5cm]{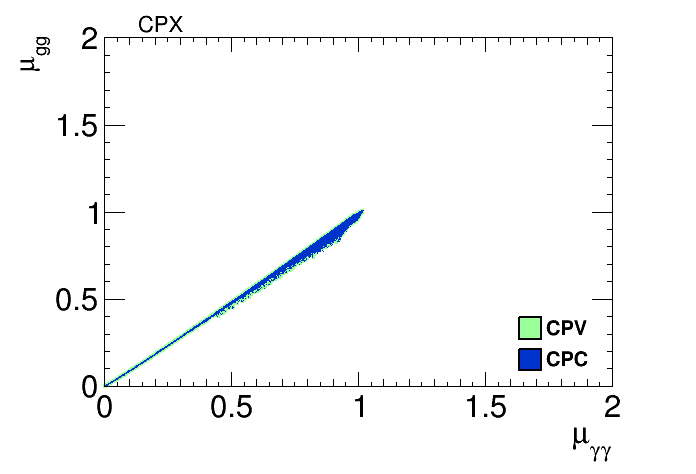}\includegraphics[width=7.5cm]{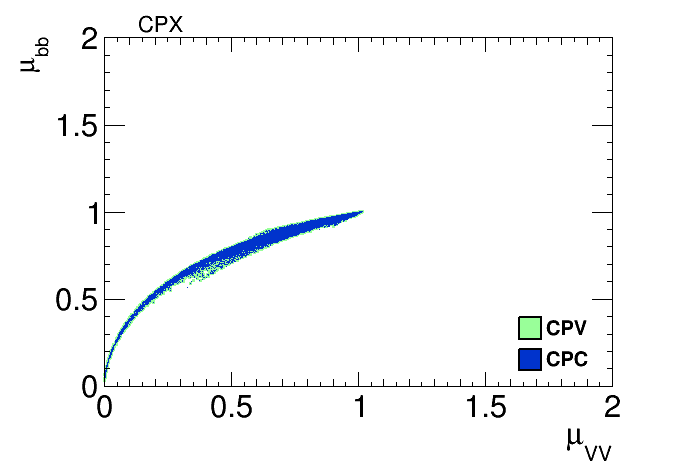}
\caption{\it Scatter plots of the $h_1$  signal strengths in the CPX scenario in the CP-violating limit
$\Phi_\alpha = 0$ (blue dots) and in the CP-violating sample (green dots).
The left panel displays a strong linear correlation between $\mu_{\gamma \gamma}$
and $\mu_{gg}$, and the right panel displays a non-linear
correlation between $\mu_{VV}$ and $\mu_{{\bar b}b}$.\label{fig:cpx-muhiggs}}
 \end{center}
\end{figure}

Fig.~\ref{fig:cpx-muhiggs} shows scatter plots of $h_1$ signal strengths $\mu_X$ in the CPX
scenario with the CP-violating phases $\Phi_\alpha =0$ (blue dots) and $\ne 0$ (green dots):
again only very small differences are seen.
In the left panel we see a strong, almost linear correlation between $\mu_{\gamma \gamma}$ and $\mu_{gg}$,
and in the right panel we see a nonlinear correlation between $\mu_{VV}$ and $\mu_{{\bar b}b}$.

As already mentioned, our results for $A_{CP}$ in the CPX scenario are very small, so we do not display them.
Taken together with the results shown in Figs.~\ref{fig:cpx-Rhiggs} and \ref{fig:cpx-muhiggs},
where no distinctive signatures of non-zero phases $\Phi_\alpha \ne 0$ are visible,
our results suggest that one should look elsewhere for probes of CP violation in the CPX scenario.

\begin{figure}[!t]
 \begin{center}
 \includegraphics[width=7.5cm]{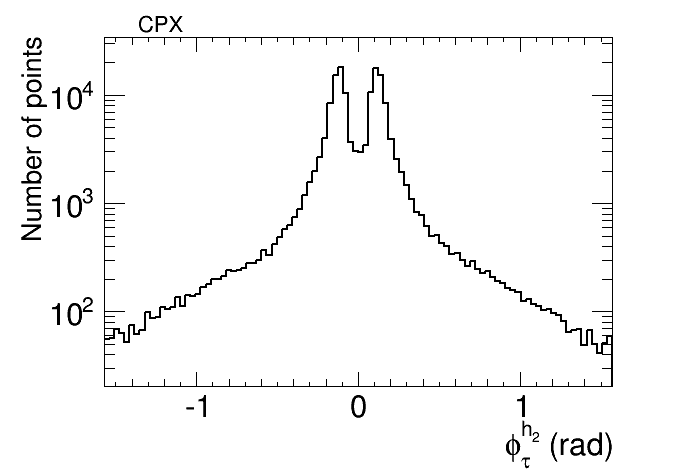}\includegraphics[width=7.5cm]{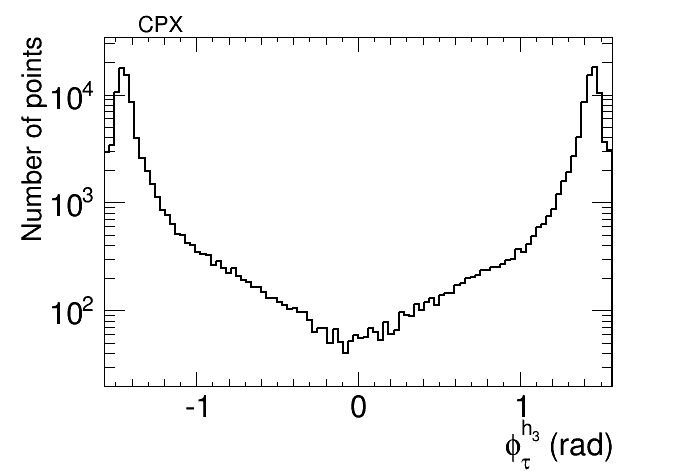}
\caption{\it The distributions of (left) the CP-violating phase $\phi_\tau^{h_2}$ in $h_2 \tau \tau$
couplings and (right) the CP-violating phase $\phi_\tau^{h_3}$ in $h_3 \tau \tau$
couplings in the CPX scenario (\protect\ref{CPX}, \protect\ref{varyCPX}),
found after  applying all the constraints using the geometric approach
described in the text.\label{fig:CPXtau}}
 \end{center}
\end{figure}

\begin{figure}[!t]
 \begin{center}
 \includegraphics[width=7.5cm]{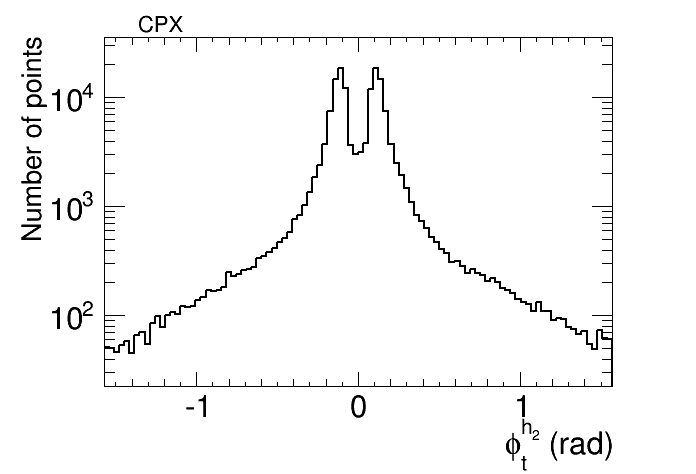}\includegraphics[width=7.5cm]{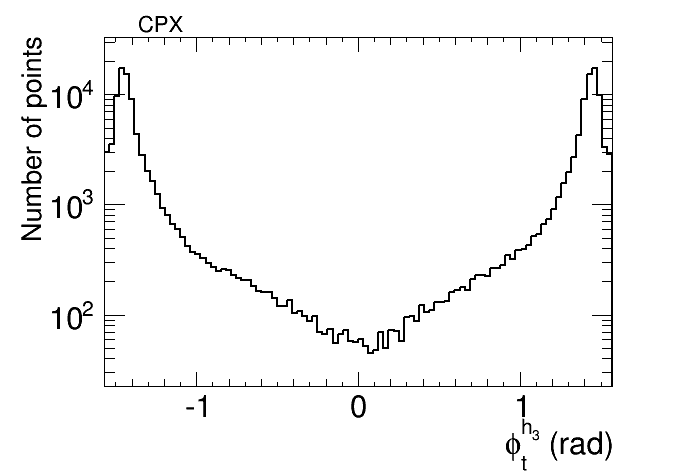}
\caption{\it The distributions of (left) the CP-violating phase $\phi_t^{h_2}$ in $h_2 {\bar t} t$
couplings and (right) the CP-violating phase $\phi_t^{h_3}$ in $h_3 {\bar t} t$
couplings in the CPX scenario (\protect\ref{CPX}, \protect\ref{varyCPX}),
found after  applying all the constraints using the geometric approach
described in the text.\label{fig:CPXt}}
 \end{center}
\end{figure}

We have also analysed the prospects for CP violation in the couplings of the neutral Higgs
bosons to $\tau^+ \tau^-$ and ${\bar t} t$ in the CPX scenario (\ref{CPX}, \ref{varyCPX}),
as given by the phases $\phi^{h_i}_\tau$ and $\phi^{h_i}_t$
for $i = 1, 2, 3$ defined in (\ref{CPVheavyH}).
As in the CMSSM case discussed previously,
we find that after imposing the EDM constraints the phases for the $h_1$ couplings are small,
$\phi^{h_i}_\tau \la 0.1$~radians and $\phi^{h_i}_t \la 0.02$~radians.
On the other hand, the phases for the $h_2$ and $h_3$ couplings may again be
quite large, as seen in Figs.~\ref{fig:CPXtau} and \ref{fig:CPXt}, respectively.
Thus $h_{2,3}$ decays may also provide interesting
prospects for probing CP violation in this CPX scenario.

\newpage
\subsection{Phenomenological MSSM (pMSSM)}
\label{sec:pmssm}

We now consider the MCPMFV version of the phenomenological MSSM (pMSSM), which has 25 parameters: the 19 real
parameters
\begin{equation}
M_{1,2,3}, \; M_{{\tilde Q_L}, {\tilde U_R}, {\tilde D_R}, {\tilde L_L}, {\tilde E_R}}, M_{{\tilde Q_{3L}}, {\tilde t_R}, {\tilde b_R}, {\tilde L_{3L}}, {\tilde \tau_R}}, \; M_{H^\pm},\mu, \tan\beta, \; A_{t, b, \tau}
\label{pMSSM}
\end{equation}
and the six phases $\Phi_\alpha$ discussed previously. We perform a scan of the pMSSM parameter space
using the iterative geometric approach described in Sec.~\ref{sec:method}. 
We first generated about 40 million points, and then kept only points with
a neutral Higgs boson with a mass in the range 121-129 GeV (thereby allowing for a conservative theoretical
uncertainty in the Higgs mass calculation), and with a neutralino LSP. These requirements
reduced the number of points to about 1 million. Imposing the EDM constraints then left about 150000 valid points.
In the following plots, in addition to these constraints, we also impose flavour constraints, the cosmological upper bound
on the dark matter density, the LUX direct upper limit on spin-independent dark matter
scattering (except when the same observable is plotted), and we require squarks and the gluino to have masses above 500 GeV.

\begin{figure}[!p]
 \begin{center}
  \includegraphics[width=7.5cm]{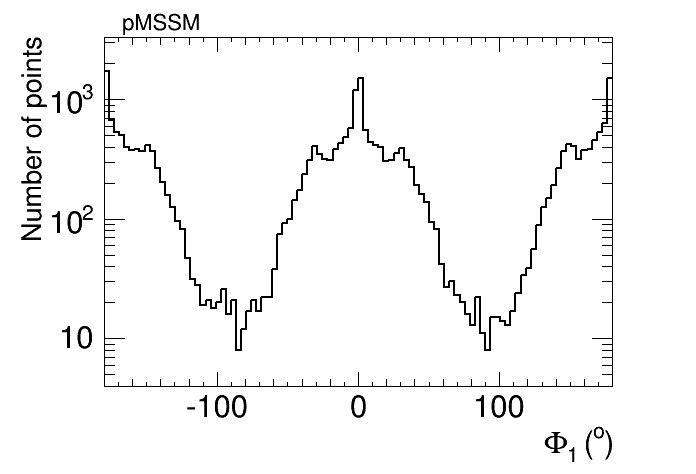}\includegraphics[width=7.5cm]{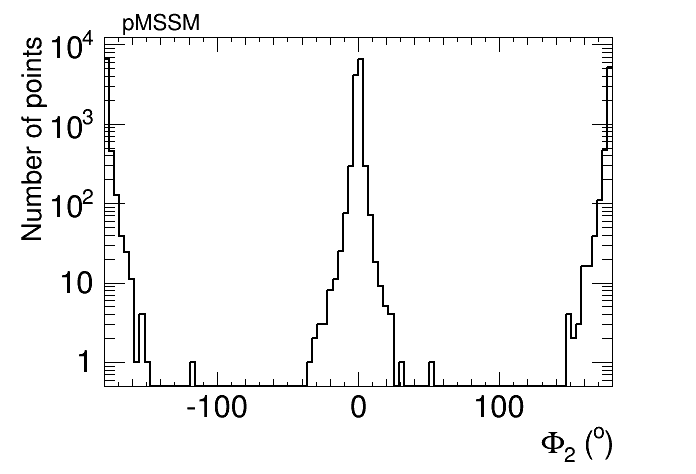}\\[0.5cm]
  \includegraphics[width=7.5cm]{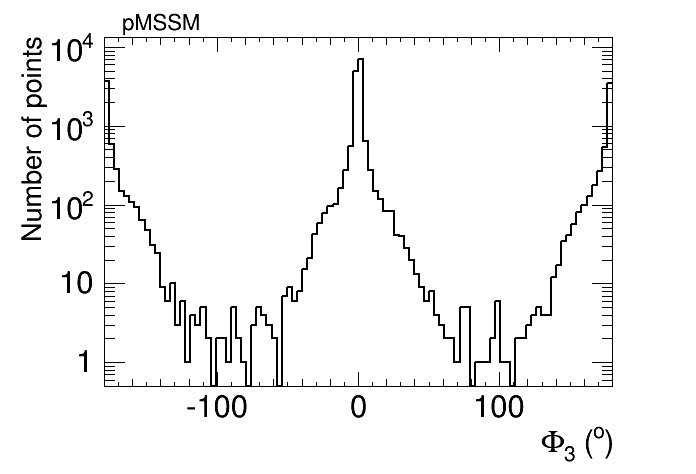}\includegraphics[width=7.5cm]{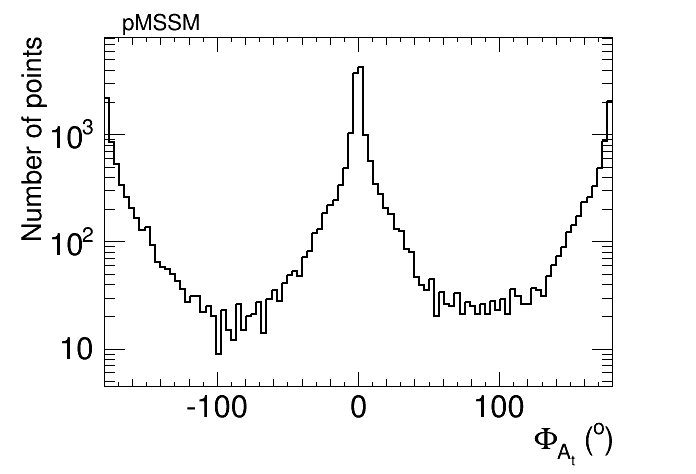}\\[0.5cm]
  \includegraphics[width=7.5cm]{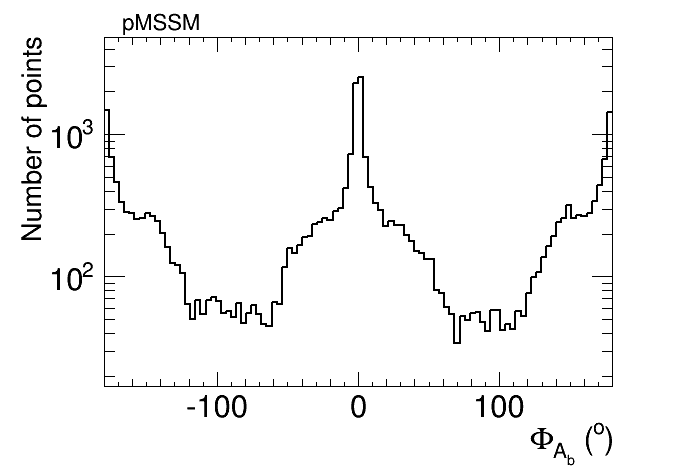}\includegraphics[width=7.5cm]{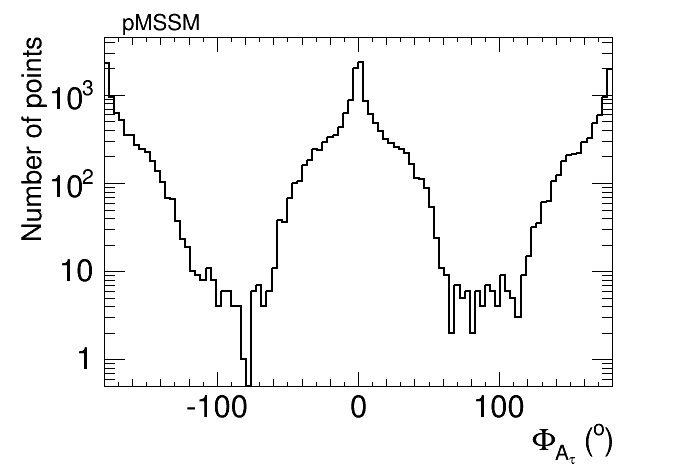}
\caption{\it Sampling of the CP-violating phases $\Phi_\alpha$  in the pMSSM scenario generated in the iterative
geometric approach, imposing the EDM and other constraints.\label{fig:distrib}}
 \end{center}
\end{figure}

Fig.~\ref{fig:distrib} shows the samplings of the phases $\Phi_\alpha$
obtained after imposing these constraints. We see that values of $\Phi_{A_{t,b}}$
and $\Phi_1 \sim \pm \ 90^\circ$ are quite well sampled, as are values of
$\Phi_{A_\tau} \sim 90^\circ$. On the other hand, large values of $\Phi_3$ are
less well sampled, and the range of $\Phi_2$ is very restricted
with only small deviations from the CP-conserving cases being allowed. 

\begin{figure}[!t]
 \begin{center}
 \includegraphics[width=7.5cm]{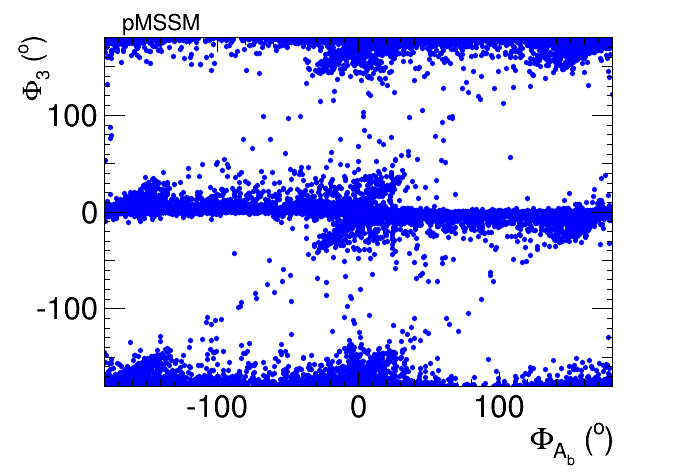}\includegraphics[width=7.5cm]{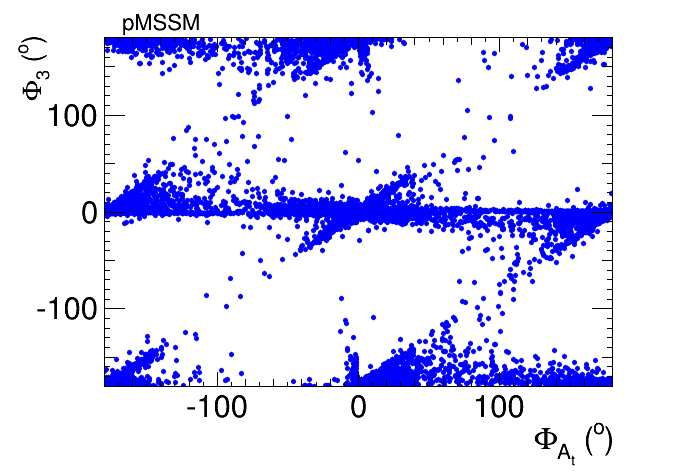}
\caption{\it Correlations of $\Phi_3$ with $\Phi_{\rm{A_b}}$ (left panel) and
$\Phi_3$ with $\Phi_{\rm{A_t}}$ (right panel) imposed by the EDM constraints in the pMSSM scenario.}
\label{fig:pMSSMcancel}
 \end{center}
\end{figure}

We see in Fig.~\ref{fig:pMSSMcancel} the extent to which the EDM constraints impose
cancellations $\Phi_3$ and $\Phi_{A_t}$ (left panel) and between $\Phi_3$ and $\Phi_{A_b}$ (right panel).
We see that large values of $(\Phi_{A_t, A_b}, \Phi_3) \sim ( \pm \ 90^\circ, \pm \ 90^\circ)$
are allowed, and we also see diagonal features corresponding to correlations.
As in the NUHM2, it is apparent that the EDM constraints do not require all the CP-violating
phases to be small simultaneously.

\begin{figure}[!t]
 \begin{center}
  \includegraphics[width=7.5cm]{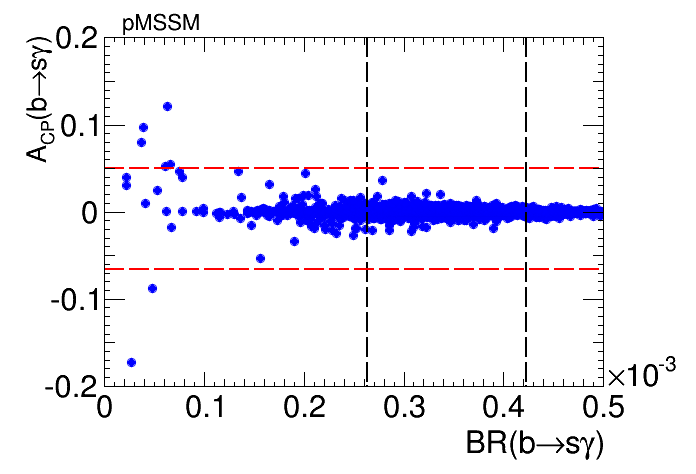}\includegraphics[width=7.5cm]{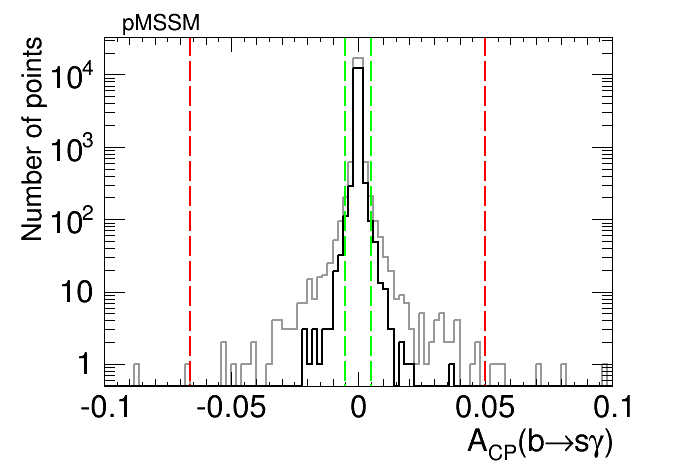}
\caption{\it Left panel: Scatter plot of the branching ratio for $b\to s\gamma$ decay versus its
CP-violating asymmetry, $A_{CP}$, in the pMSSM scenario. The vertical black dashed lines represent
the allowed range for the $b \to s \gamma$ branching ratio, and the horizontal red dashed lines
represent the present experimental limits on $A_{CP}$. Right panel: Histogram
of $A_{CP}$ in the pMSSM, imposing only the Higgs mass and EDM cuts (grey: full sample, black:
points satisfying the EDM constraints). The vertical red
dashed lines represent the present experimental limits, and the vertical green dashed lines
represent the prospective future improvement in the sensitivity to $A_{CP}$ by a factor of 10.\label{fig:pMSSMACPbsg}}
 \end{center}
\end{figure}

The left panel of Fig.~\ref{fig:pMSSMACPbsg} displays a scatter plot of the values of $A_{CP}$ found in the pMSSM
using the iterative geometric approach. We see that values $\la 3$\% are possible for
values of the $b \to s \gamma$ branching ratio lying within the experimentally allowed range.
The right panel of Fig.~\ref{fig:pMSSMACPbsg} shows a histogram of $A_{CP}$ values,
imposing only the Higgs mass and EDM cuts. Here we see tails extending to larger values of
$|A_{CP}|$ that lie outside the experimentally allowed range when the EDM constraints are not applied
(grey histogram), whereas the
black histogram is for points satisfying the EDM constraints.
The vertical red dashed lines show the present experimental constraints
on $A_{CP}$, and the possible future improvement in the experimental sensitivity by a factor of 10
is indicated by vertical green dashed lines. As in the NUHM2, there are CP-violating
pMSSM parameter sets that could be explored with such an improvement: it would provide additional information on CP violation
within the pMSSM.

\begin{figure}[!t]
 \begin{center}
   \includegraphics[width=8.5cm]{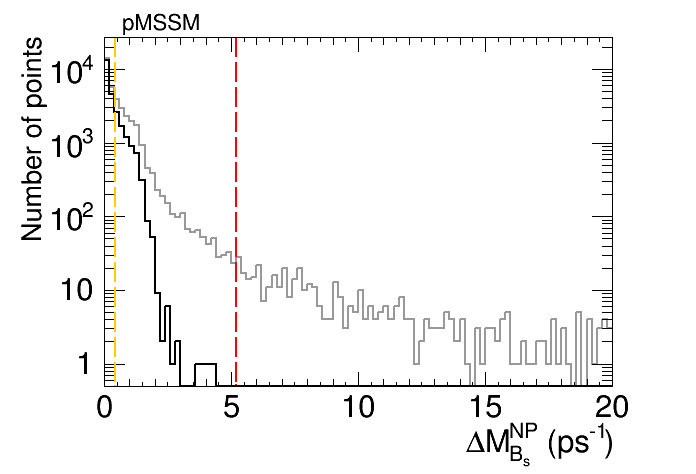}
\caption{\it Histogram of the possible new physics contribution to $B_s$ mixing, $\Delta M^{NP}_{B_s}$,
in the pMSSM scenario. The grey histogram is for points satisfying the Higgs mass and other constraints,
and the black histogram is for points that also satisfy the EDM constraints. The vertical red dashed
line is the present experimental upper limit on $\Delta M^{NP}_{B_s}$, and the vertical yellow dashed line
shows the potential of a reduction in the current theoretical uncertainty in the Standard Model by a factor of 10.
\label{fig:pMSSMBs}}
 \end{center}
\end{figure}

The possible new physics contribution to $B_s$ meson
mass mixing, $\Delta M^{NP}_{B_s}$, in the pMSSM scenario is shown in Fig.~\ref{fig:pMSSMBs}.
As in the previous cases studied, the grey
histogram is for the full sample, and the black histogram is for
points that also satisfy the EDM constraints. If the theoretical uncertainty in the Standard Model contribution to $B_s$ mixing
could be reduced by a factor of 10 thanks to improved lattice calculations, the sensitivity to
$\Delta M^{NP}_{B_s}$ would become that indicated by the vertical yellow dashed line in Fig.~\ref{fig:pMSSMBs}.
In this case, many of the pMSSM models that are currently viable (indicated by the black histogram) could be explored.

\begin{figure}[!t]
 \begin{center}
  \includegraphics[width=7.5cm]{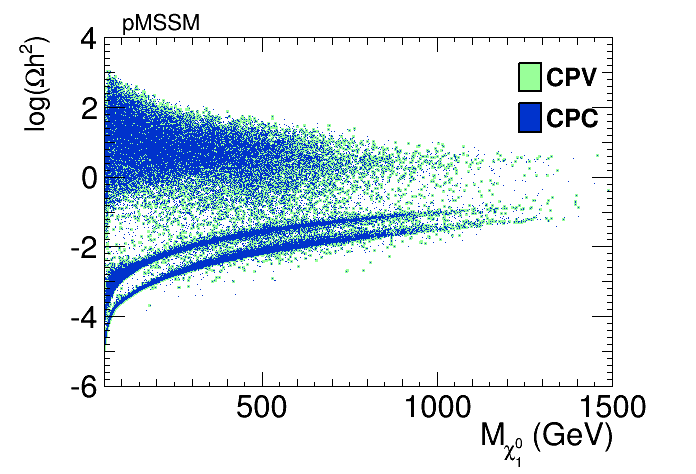}  \includegraphics[width=7.5cm]{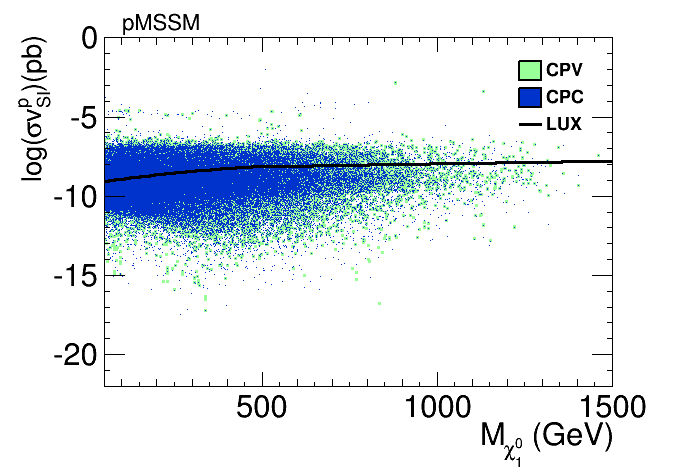}
\caption{\it Left panel: Scatter plot of the dark matter relic density as a function of the neutralino mass in the pMSSM scenario.
Right panel: Scatter plot of the spin-independent dark matter scattering cross section
$\sigma_{SI}^p v$ as a function of the neutralino mass in the pMSSM scenario.
In both panels, CP-conserving parameter choices are denoted by blue dots, and CP-violating parameter
choices by green dots.\label{fig:oh2}}
 \end{center}
\end{figure}

In Fig.~\ref{fig:oh2}, we show in the left panel
the values of the relic LSP density $\Omega_\chi h^2$ that we find in our pMSSM scan,
and in right panel we show values of the spin-independent dark matter scattering
cross section $\sigma_{SI}^p v$. We see that values of $\Omega_\chi h^2$ considerably
above the cosmological upper limit are possible in both the CP-conserving (blue dots)
and CP-violating cases (green dots).  We also see in the right panel of Fig.~\ref{fig:oh2}
that values of $\sigma_{SI}^p v$ above the LUX upper limit are also possible. In both panels,
there are no large differences between the CP-conserving and CP-violating cases.

\begin{figure}[!t]
\vspace*{-0.3cm} \begin{center}
  \includegraphics[width=7.5cm]{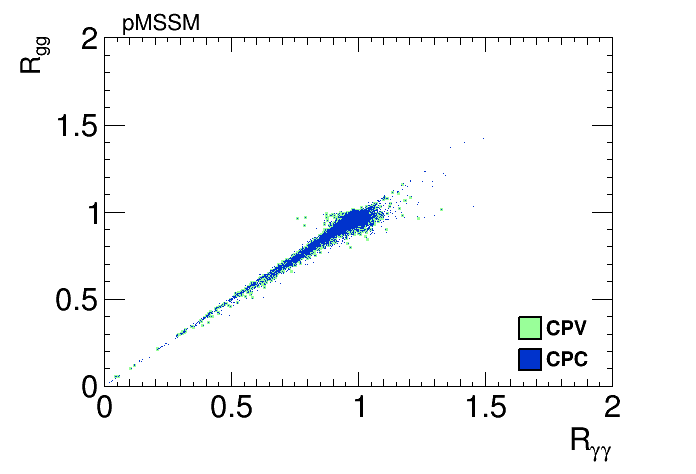}\includegraphics[width=7.5cm]{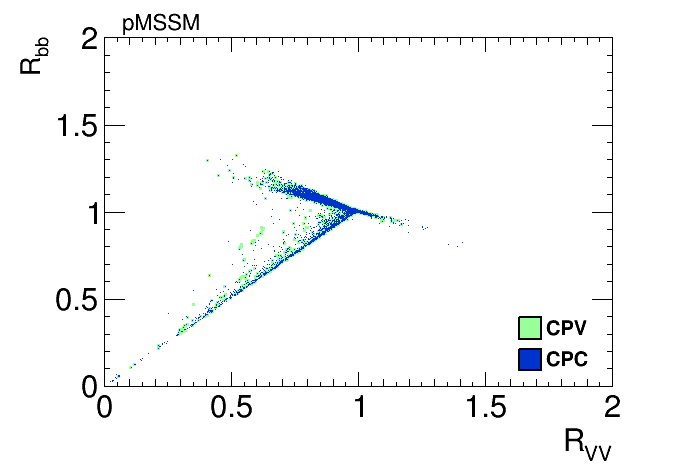}
\caption{\it Scatter plots of branching ratios, normalised to the Standard Model values,
for decays of the lightest Higgs boson, $h_1$,
in the pMSSM scenario in the CP-violating limit $\Phi_\alpha = 0$ (blue dots) and in the CP-violating
sample (green dots). The left panel displays a strong correlation between $R_{\gamma \gamma}$
and $R_{gg}$, and the right panel displays a bimodal correlation between $R_{VV}$ and $R_{{\bar b}b}$.
\label{fig:Rhiggs}}
 \end{center}
 \vspace*{-0.3cm} 
\end{figure}

Scatter plots of values of $h_1$ branching ratios  in the pMSSM scenario are in Fig.~\ref{fig:Rhiggs},
the left panel displaying $(R_{\gamma \gamma}, R_{gg})$ and the right panel displaying $(R_{VV}, R_{{\bar b}b})$.
As previously, the blue dots are CP-conserving parameter choices with $\Phi_\alpha = 0$, and the green dots
are from a scan of CP-violating points with $\Phi_\alpha \ne 0$. 

As in the NUHM2 scenario,
we note in the left panel a strong correlation between
$R_{\gamma \gamma}$ and $R_{gg}$, which may be either much smaller than in the Standard Model or somewhat larger,
and we also see in the right panel that a large reduction in $R_{VV}$ is possible. Also as in the NUHM2 scenario,
the reduction in $R_{VV}$ may be accompanied by values of $R_{{\bar b}b}$ that are either larger or smaller than
in the Standard Model, the latter possibility arising when the Higgs boson can decay into light sparticles.

\begin{figure}[!t]
 \begin{center}
\includegraphics[width=7.5cm]{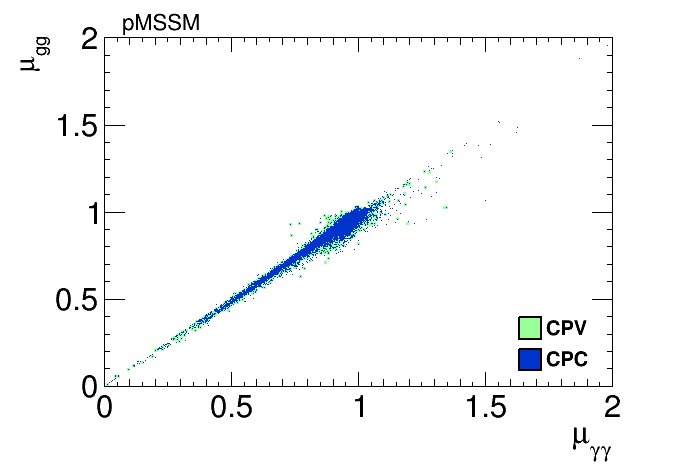}\includegraphics[width=7.5cm]{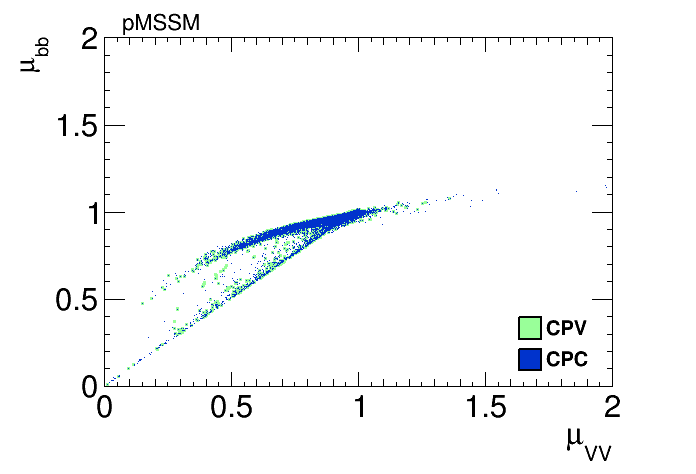}
\caption{\it Scatter plots of the $h_1$  signal strengths in the pMSSM scenario in the CP-violating limit
$\Phi_\alpha = 0$ (blue dots) and in the CP-violating sample (green dots).
The left panel displays a strong linear correlation between $\mu_{\gamma \gamma}$
and $\mu_{gg}$, and the right panel displays a bimodal
correlation between $\mu_{VV}$ and $\mu_{{\bar b}b}$ for smaller values.\label{fig:muhiggs}}
 \end{center}
\end{figure}

Fig.~\ref{fig:muhiggs} displays scatter plots of $h_1$ signal strengths $\mu_X$ in the pMSSM
scenario in the CP-conserving case with phases $\Phi_\alpha =0$ (blue dots) and
in the CP-violating case where the $\Phi_\alpha \ne 0$ (green dots).
As in the NUHM2 case, we see a strong correlation between $\mu_{\gamma \gamma}$ and $\mu_{gg}$ in the left panel,
and in the right panel we see a correlation between $\mu_{VV}$ and $\mu_{{\bar b}b}$ that becomes bimodal
for small values of $\mu_{VV}$.

\begin{figure}[!t]
 \begin{center}
  \includegraphics[width=7.5cm]{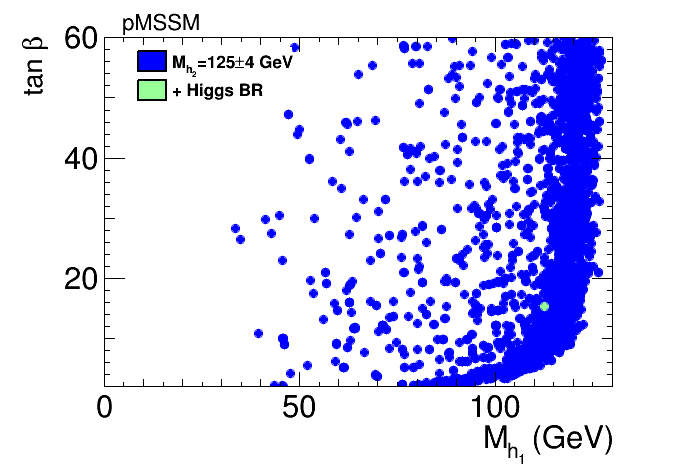}\includegraphics[width=7.5cm]{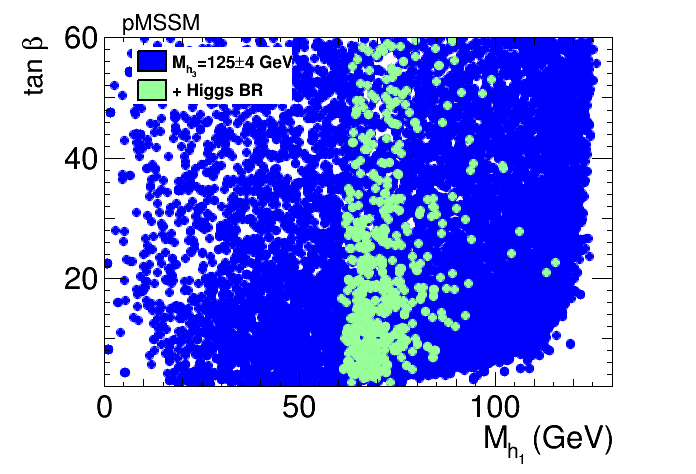}\vspace*{-0.3cm}
\caption{\it Scatter plots of pMSSM points in the ($m_{h_1},\tan\beta$) plane in the case where either the $h_2$ (left panel)
or the $h_3$ (right panel) is the Higgs boson discovered at the LHC, applying only the
EDM and Higgs mass constraints (blue dots), and applying also the Higgs signal strength constraints (green dots).
We find no points that satisfy in addition the neutral and charged heavy Higgs search constraints.
\label{fig:heavyhiggs}\vspace*{-0.3cm}}
 \end{center}
\end{figure}

\begin{figure}[!t]
 \begin{center}
  \includegraphics[width=7.5cm]{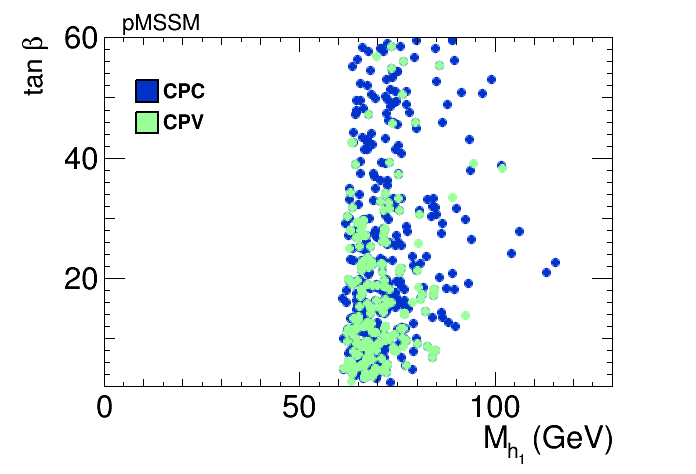}\vspace*{-0.3cm}
\caption{\it Scatter plot of pMSSM points in the ($m_{h_1},\tan\beta$) plane in the case where the Higgs boson
discovered at the LHC is identified as the $h_3$, for the points satisfying the Higgs signal strength constraints
as well as the EDM constraints.\label{fig:heavyhiggs2}\vspace*{-0.3cm}}
 \end{center}
\end{figure}

We have also studied whether the Higgs boson discovered at the LHC might be one of the heavier Higgs bosons
in the pMSSM, with or without CP violation. As seen in the left panel of Fig.~\ref{fig:heavyhiggs}, if the known
Higgs boson is identified with the $h_2$, it is not possible to satisfy the Higgs signal strength constraints.
This is possible if the discovered Higgs boson is identified with the $h_3$, as seen (green dots) in the right panel
of Fig.~\ref{fig:heavyhiggs}, in which case the $h_1$ mass is about 60 - 80 GeV. Fig.~\ref{fig:heavyhiggs2}
displays these points in both the CP-conserving case (blue dots) and the CP-violating case (green dots),
which are quite similar. On the other hand, none of these points
survive the charged Higgs and $A/H \to \tau\tau$ constraints, nor the flavour constraints.
We therefore conclude that the pMSSM does not provide a way to conceal a neutral Higgs boson
that is lighter than the one discovered, even if CP is violated.

\begin{figure}[!t]
 \begin{center}
 \includegraphics[width=7.5cm]{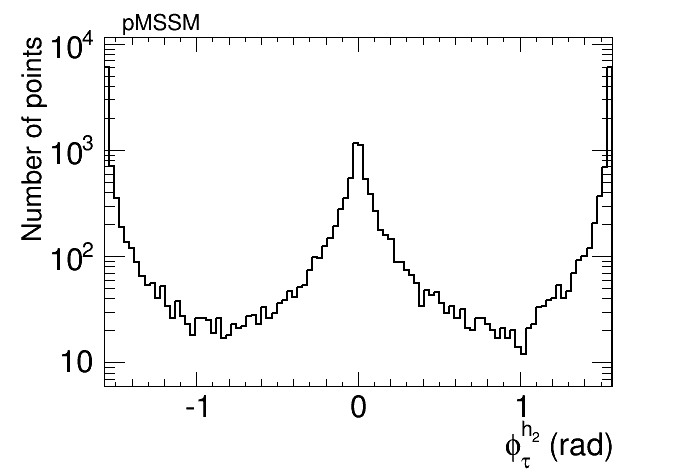}
 \includegraphics[width=7.5cm]{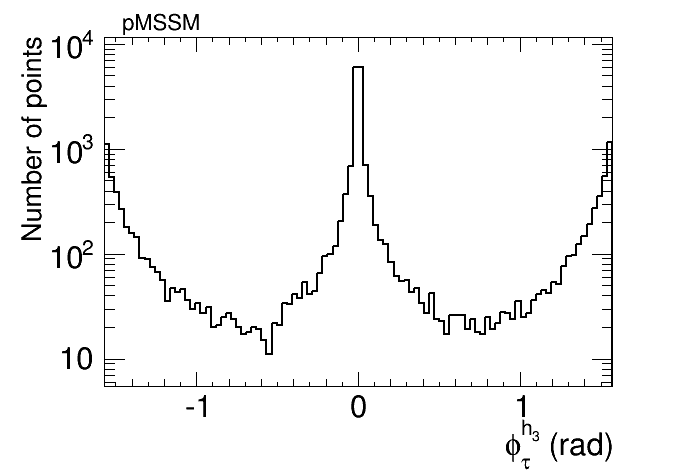}
\caption{\it The distributions of (left) the CP-violating phase $\phi_\tau^{h_2}$ in $h_2 \tau \tau$
couplings and (right) the CP-violating phase $\phi_\tau^{h_3}$ in $h_3 \tau \tau$
couplings in the pMSSM scenario (\protect\ref{pMSSM}),
found after  applying all the constraints using the geometric approach
described in the text.\label{fig:pMSSMtau}}
 \end{center}
\end{figure}

\begin{figure}[!t]
 \begin{center}
 \includegraphics[width=7.5cm]{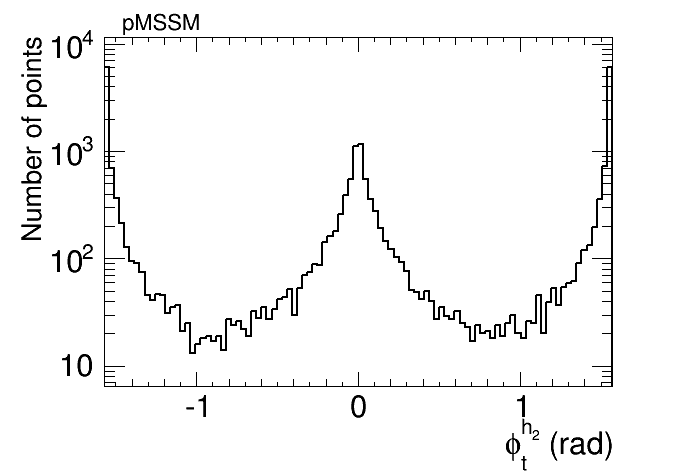}
 \includegraphics[width=7.5cm]{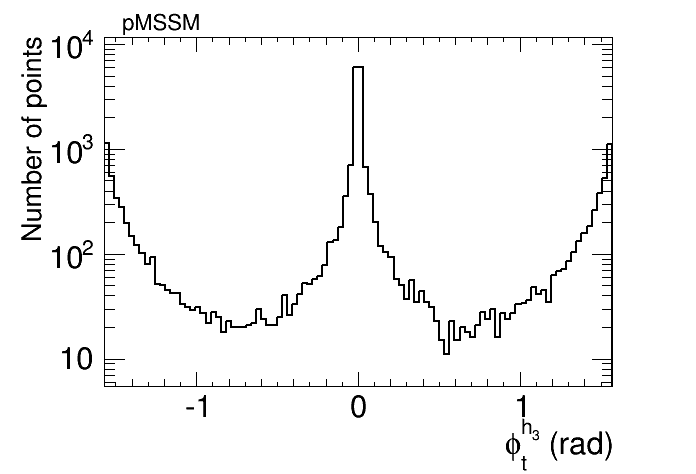}
\caption{\it The distributions of (left) the CP-violating phase $\phi_t^{h_2}$ in $h_2 {\bar t} t$
couplings and (right) the CP-violating phase $\phi_t^{h_3}$ in $h_3 {\bar t} t$
couplings in the pMSSM scenario (\protect\ref{pMSSM}),
found after  applying all the constraints using the geometric approach
described in the text.\label{fig:pMSSMt}}
 \end{center}
\end{figure}

Assuming that the Higgs boson discovered at the LHC is indeed the lightest MSSM Higgs boson $h_1$,
we now assess the prospects for CP violation in the couplings of the heavy neutral Higgs
bosons to $\tau^+ \tau^-$ and ${\bar t} t$ in the pMSSM scenario (\ref{pMSSM})
which are shown in Figs.~\ref{fig:pMSSMtau} and \ref{fig:pMSSMt}.
We see that, as in the CMSSM, CPX and NUHM2 cases discussed previously,
$h_{2,3}$ decays may provide interesting
prospects for probing CP violation also in this pMSSM scenario.
On the other hand, we again find that after imposing all the constraints the phases for the $h_1$ couplings are small,
namely $\phi^{h_1}_\tau \la 0.03$~radians and $\phi^{h_1}_t \la 0.02$~radians, respectively.

\section{Conclusions}

The geometrical approach to implementing EDM constraints and maximising other CP-violating
observables proposed in~\cite{Ellis:2010xm} provides a suitable way to explore the possibilities
for CP violation in variants of the MSSM, which we have applied in this paper to explore the CMSSM,
the CPX scenario, the NUHM2 and the pMSSM. We have adopted an iterative extension of the
geometric approach, which is suitable for exploring larger values of the CP-violating phases.
Our explorations have been within the
maximally CP-violating, minimal flavour-violating (MCPMFV) framework with six CP-violating phases,
of which two combinations are unconstrained {\it a priori} by the four EDM constraints. The following
are our principal results:

$\bullet$ In the CMSSM we have explored CP-violating generalisations of the low-mass best-fit point
(\ref{CMSSMBF}) that was identified in~\cite{Buchmueller:2013rsa,Buchmueller:2014yva}, where we found relatively little scope
for large deviations from the CP-conserving case, e.g., in the masses of the Higgs bosons and the
spin-independent dark matter scattering cross section. Moreover, we found that only very small
values of $A_{CP} \la 0.001$ would be possible in this case, and the new physics contribution to
$B_s$ meson mixing, $\Delta M^{NP}_{B_s}$, would not be observable.

$\bullet$ We have then explored the CPX scenario (\ref{CPX}), where we also found no scope for
measurable values of $A_{CP}$. On the other hand, we found in this model that 
$\Delta M^{NP}_{B_s}$ could be large enough
to provide a possible signature if the current lattice theoretical uncertainty in the Standard Model
contribution to $B_s$ mixing could be reduced by a factor of 10, as seen in Fig.~\ref{fig:CPXBs}.

$\bullet$ The situation in the NUHM2 scenario (\ref{NUHM2ranges}) is rather more favourable for
observable signals of CP violation. In this case, $A_{CP}$ could be as large as $\sim 2$\%
and hence lie well within the reach
of experiment, as seen in Fig.~\ref{fig:nuhm-ACPbsg}, and $\Delta M^{NP}_{B_s}$
might also be large enough to provide a possible experimental signature, as seen in
Fig.~\ref{fig:NUHM2Bs}.

$\bullet$ A similar situation was found in the pMSSM scenario (\ref{pMSSM}), in which case
$A_{CP}$ could be as large as $\sim 3$\%, as seen in Fig.~\ref{fig:pMSSMACPbsg},
again within the reach of experiment. We also find in this scenario that $\Delta M^{NP}_{B_s}$
could be large enough to be observable with a prospective reduction in the theoretical
uncertainty in the Standard Model calculation of $B_s$ mixing, as seen in Fig.~\ref{fig:pMSSMBs}.

$\bullet$ In all the scenarios studied, the CP-violating phases in the $h_1 \tau^+ \tau^-$
and $h_1 {\bar t} t$ couplings are small. On the other hand, the phases in in the
$h_{2,3} \tau^+ \tau^-$ and $h_{2,3} {\bar t} t$ couplings can be quite large, and may
present interesting prospects for future $pp$, $e^+ e^-$ and $\mu^+ \mu^-$ experiments,
though their detailed study lies beyond the scope of this work.

Our analysis serves as a reminder that the EDM constraints do not force all the six non-KM
CP-violating phases in MCPMFV to be small, and that in some variants of the MSSM
there could be observable signatures of CP violation beyond the Standard Model, e.g., $A_{CP}$ in
$b \to s \gamma$ decay. We look forward to a generation of $A_{CP}$ measurements, and also
to improved theoretical calculations of the Standard Model contribution to $B_s$ meson mixing,
which might enable a new physics contribution $\Delta M^{NP}_{B_s}$ to be isolated.
If enough soft supersymmetry-breaking parameters could be measured,
and both $A_{CP}$ and $\Delta M^{NP}_{B_s}$ could be shown to have measurable deviations
from the Standard Model, one might finally be able to fix all the six non-KM
CP-violating phases in MCPMFV.

\vspace*{-0.1cm}
\section*{Acknowledgements}

The work of J.E. was supported in part by the London Centre for Terauniverse Studies
(LCTS), using funding from the European Research Council via the Advanced Investigator
Grant 267352 and from the UK STFC via the research grant ST/J002798/1. The work
of A.A. was supported in part by the F\'ed\'eration de Recherche A.-M. Amp\`ere de Lyon. R.M.G. wishes to acknowledge support from the Department of Science and Technology, India under Grant No. SR/S2/JCB-64/2007 under the J.C. Bose  Fellowship scheme and hospitality in the CERN theory division.


\end{document}